\documentclass[epj,nopacs]{svjour}
\usepackage{epsfig}
\usepackage{latexsym}
\usepackage{amssymb}

 \newcommand{\be}{\begin{equation}}
 \newcommand{\ee}{\end{equation}}
 \newcommand{\bea}{\begin{eqnarray}}
 \newcommand{\eea}{\end{eqnarray}}
 \newcommand{\nn}{\nonumber}

 \journalname{EPJ C}
 
\begin{document}
 \setcounter{page}{1}
\title{
Complete description of polarization effects in emission of\\ a
photon by an electron in the field of a strong laser wave}
\titlerunning{Complete description of polarization effects in
emission of a photon by an electron}
\authorrunning{D.Yu.~Ivanov et al.}
\author{D.Yu.~Ivanov \inst{1,}\thanks{d-ivanov@math.nsc.ru} %
\and G.L.~Kotkin \inst{2,}\thanks{kotkin@math.nsc.ru} %
\and V.G.~Serbo \inst{2,}\thanks{serbo@math.nsc.ru}}
\institute{Sobolev Institute of Mathematics, Novosibirsk, 6300090,
Russia \and Novosibirsk State University, Novosibirsk, 630090
Russia}
\date{Received: February 13, 2004}

\abstract{
 We consider emission of a photon by an electron in the
field of a strong laser wave. Polarization effects in this process
are important for a number of physical problems. The probability
of this process for circularly or linearly polarized laser photons
and for arbitrary polarization of all other particles is
calculated. We obtain the complete set of functions which describe
such a probability in a compact invariant form. Besides this, we
discuss in some detail the polarization effects in the kinematics
relevant to the problem of $e\to \gamma$ conversion at $\gamma
\gamma$ and $\gamma e$ colliders.
 }

 \maketitle


\section{Introduction}

The Compton scattering
 \be
  e(p) +\gamma (k) \to e(p') +\gamma (k')
  \label{1}
 \ee
is one of the first processes calculated in quantum
electrodynamics at the end of 1920's. The analysis of polarization
effects in this reaction is now included in text-books (see, for
example, \cite{BLP} \S 87). Nevertheless, the complete description
of cross sections with polarization of both initial and final
particles has been considered in detail only recently (see
\cite{Grozin,GKPS83,KPS98} and literature therein). One
interesting application is the collision of an ultra-relativistic
electron with a beam of polarized laser photons. In this case the
Compton effect is the basic process for obtaining of high-energy
photons for contemporary experiments in nuclear physics
(photo-nuclear reactions with polarized photons) and for future
$\gamma \gamma $ and $\gamma e$ colliders \cite{GKST83}. The
importance of the particle polarization is clearly seen from the
fact that in comparison with the unpolarized case the number of
final photons with maximum energy is nearly doubled when the
helicities of the initial electron and photon are
opposite~\cite{GKPS83}.

With the growth of the laser field intensity, an electron starts
to interact coherently with $n$ laser photons,
 \be
   e(q) +n\,\gamma_L (k) \to e(q') +\gamma (k')\,,
 \label{2}
 \ee
thus the Compton scattering becomes non-linear. Such a process
with absorption of $n=1,\, 2,\, 3,\, 4$ linearly polarized laser
photons was observed in the recent experiment at SLAC~\cite{SLAC}.
The polarization properties of the process (\ref{2}) are important
for a number of problems, for example, for laser beam
cooling~\cite{Telnov} and, especially, for future $\gamma \gamma $
and $\gamma e$ colliders (see~\cite{GKPnQED,GKLT2001} and the
literature therein) . In the latter case the non-linear Compton
scattering must be taken into account in simulations of the
processes in the conversion region. For comprehensive simulation,
including processes of multiple electron scattering, one has to
know not only the differential cross section of the non-linear
Compton scattering with a given number of the absorbed laser
photons $n$, but energy, angles and polarization of final photons
and electrons as well. The method of calculation for such cross
sections was developed by Nikishov and Ritus~\cite{NR-review}. It
is based on the exact solution of the Dirac equation in the field
of the external electromagnetic  plane wave. Some particular
polarization properties of this process were considered
in~\cite{NR-review}--\cite{BPM} and have already been included in
the existing simulation codes~\cite{Yokoya,TCode}. Another
approach was used in~\cite{BKMS}, where the total cross section
and the spectrum for the non-linear Compton scattering, summed
over numbers of the absorbed laser photons $n$ and over spin
states of the final particles, had been obtained.

In the present paper we give the complete description of the
non-linear Compton scattering for the case of circularly or
linearly polarized laser photons and arbitrary polarization of all
other particles\footnote{Preliminary results of this work have
been submitted to the hep-ph archive \cite{IKS}.}. We follow the
method of Nikishov and Ritus in the form presented in~\cite{BLP}
\S 101. In the next section we describe  in detail the kinematics
of the process ({\ref{2}). The effective differential cross
section is obtained in Sect. 3 in the compact invariant form,
including the polarization of all particles. In Sect. 4 we
consider the process (\ref{2}) in the reference frame relevant for
$\gamma \gamma$ and $\gamma e$ colliders. The limiting cases are
discussed in Sect. 5. In Sect. 6 we present some numerical results
obtained for the range of parameters close to those in the
existing projects of the $\gamma \gamma$ and $\gamma e$ colliders.
In the last section we summarize our results and compare them with
those known in the literature. In Appendix A we show that our
results coincide in the limit of weak laser field with those for
the linear Compton scattering.

\section{Kinematics}

\subsection{Parameter of non-linearity}

Let us consider the interaction of an electron with a
mono\-chro\-matic plane wave described  by the 4-potential
$A_\mu(x)$. The corresponding electric and magnetic fields are
${\bf E}$ and ${\bf B}$, a frequency is $\omega$, and let $F$ be
the root-mean-squared field strength,
 $$
 F^2 =\langle {\bf B}^2 \rangle = \langle {\bf E}^2 \rangle\,.
  $$
The invariant parameter describing the intensity of the laser
field (the parameter of non-linearity) is defined via the mean
value of the squared 4-potential:
 \be
\xi={e\over mc^2} \sqrt{-\langle A_\mu (x) A^\mu(x) \rangle}\,,
 \label{3}
 \ee
where $e$ and $m$ are the electron charge and the mass, and $c$ is
the velocity of light. We use this definition of $\xi$ both for
the circularly and linearly polarized laser photons\footnote{Note
that our definition of this parameter for the case of the linear
polarization differs from the one used in~\cite{NR-review} by a
factor $1/\sqrt{2}$.}.

The origin of this parameter can be explained as follows. The
electron oscillates in the transverse direction under the
influence of a force $\sim eF$ and for the time $\sim 1/ \omega$
acquires the transverse momentum $\sim p_\perp= eF / \omega$, thus
for the longitudinal motion the effective electron mass is $m_*=
\sqrt{m^2+(p_\perp/c)^2}$. The ratio of the momentum $p_\perp$ to
$mc$ is the natural dimensionless parameter
 \be
 \xi = {eF\over m\,c\,\omega}\,.
 \label{4}
  \ee
This parameter can be expressed also via the density $n_{\rm L}$
of photons in the laser wave:
 \be
\xi^2 = \left({e F \over m \omega c}\right)^2={4\pi \alpha
\hbar^2\over m^2c\,\omega}\, n_{\rm L}\,,
 \label{5}
  \ee
where $\hbar$ is the Plank constant and $\alpha = e^2/(\hbar c)
\approx 1/137$.

\subsection{Invariant variables}

~From the classical point of view, the oscillated electron emits
harmonics with frequencies $n\,\omega$, where $n=1,\;2,\;\dots$
Their intensities at small $\xi^2$ are proportional to $({\bf
E}^2)^n \propto \xi^{2n}$, the polarization properties of these
harmonics depend on the polarizations of the laser wave and the
initial electron\footnote{The polarization of the first harmonic
is related to the tensor of the second rank $\langle
E_iE^*_j\rangle$; in this case one needs only three Stokes
parameters. The polarization properties of higher harmonics are
connected with the tensors of higher rank $\langle E_iE_j \dots
E^*_k\rangle$, and in this case one needs more parameters for
their description. It is one of the reason why the non-linear
Compton effect has been considered so far only for 100\% polarized
laser beam, mainly for circular or linear polarization.}. From the
quantum point of view, this radiation can be described as the
non-linear Compton scattering with absorption of $n$ laser
photons. When describing such a scattering, one has to take into
account that in a laser wave the 4-momenta $p$ and $p'$ of the
free initial and final electrons are replaced by the
4-quasi-momenta $q$ and $q'$ (similar to the description of a
particle motion in a periodic potential field in non-relativistic
quantum mechanics),
 \bea
q&=&p+\xi^2{ m^2c^2\over 2pk}\,k\,,\;\; q'=p'+\xi^2 {m^2c^2\over
2p'k}\,k\,,
 \label{6}\\
 q^2&=&(q')^2= (1+\xi^2 )\,m^2c^2\equiv
m_*^2c^2\,.
 \nn
 \eea
In particular, the energy of the free incident electron $E$ is
replaced by the quasi-energy
 \be
c q_0=E+\xi^2 {m^2c^2\over 2pk}\,\hbar\omega\,.
 \label{7}
 \ee

As a result, we deal with the reaction (\ref{2}) for which the
conservation law reads
 \be
 q+n\,k=q'+k'\,.
 \label{8}
 \ee
From this it follows that all the kinematic relations which occur
for the linear Compton scattering will apply to the process
considered here if the electron momenta $p$ and $p'$ are replaced
by the quasi-momenta $q$ and $q'$ and the incident photon momentum
$k$ by the 4-vector $n k$. Since $qk=pk$, it is convenient to use
the same invariant variables  as for the linear Compton scattering
(compare \cite{GKPS83}):
 \be
x={2p k\over m^2c^2}\,, \;\;\; y\; = {k k'\over p k}\,.
 \label{9}
 \ee
Moreover, many kinematic relations can be obtained from that for
the linear Compton scattering by the replacement $\omega \to n\,
\omega$, $m\to m_*$. In particular, we introduce the auxiliary
combinations
\begin{equation}
s_n=2\sqrt{r_n(1-r_n)},\;\; c_n= 1-2r_n\,,
  \label{10}
\end{equation}
where
 \be
 r_n={y\,(1+\xi^2)\over (1-y)\,nx}\,.
 \label{11}
 \ee
It is useful to note that these invariants have a simple notion,
namely
 \be
s_n=\sin{\tilde\theta},\;\; c_n=\cos{\tilde\theta}\,,\;\;
r_n=\sin^2(\tilde \theta/2)\,,
 \label{12}
 \ee
where $\tilde\theta$ is the photon scattering angle in the frame
of reference where the initial electron is at rest on average
(${\bf q}=0, \, q_0= m_* c$). Therefore,
 \be
0 \leq s_n,\; r_n \leq 1\,;\;\;-1 \leq c_n \leq 1\,.
 \label{13}
 \ee

The maximum value of the variable $y$ for the reaction (\ref{2})
is
 \begin{equation}
 y\leq \;y_n\;= {nx\over nx+1+\xi^2}\,.
\label{14}
\end{equation}
The value of $y_n$ is close to $1$ for large $n$, but for a given
$n$ it decreases with the growth of the non-linearity parameter
$\xi^2$. With this notation one can rewrite $s_n$ in the form
 \be
  s_n= {2\over y_n(1-y)}\, \sqrt{y(y_n-y)(1-y_n)}\,,
 \label{15}
 \ee
from which it follows that
 \be
 s_n \to 0 \;\;\mbox{ at } \;\;y\to y_n \;\; \mbox{or at} \;\;
 y\to0\,.
 \label{16}
 \ee

The usual notion of the cross section is not applicable for the
reaction (\ref{2}) and usually its description is given in terms
of the probability of the process per second $\dot{W}^{(n)}$.
However, for the procedure of simulation in the conversion region
as well as for the simple comparison with the linear case, it is
useful to introduce the ``effective cross section'' given by the
definition
 \be
d\sigma^{(n)}= {d\dot{W}^{(n)}\over j}\,,
 \label{17}
 \ee
where
 $$
j= {(q\,k) c^2\over q_0\, \hbar\omega}\,n_{\rm L} ={m^2c^4
\,x\over 2q_0 \hbar\omega} \, n_{\rm L}
 $$
is the flux density of colliding particles. Contrary to the usual
cross section, this effective cross section does depend on the
laser beam intensity, i.e. on the parameter  $\xi^2$. The total
effective cross section is defined as the sum over harmonics,
corresponding to the reaction (\ref{2}) with a given number $n$ of
the absorbed laser photons\footnote{In this formula and below the
sum is over those $n$ which satisfy the condition $y< y_n$, i.e.
this sum runs from some minimal value $n_{\min}$ up to $n=\infty$,
where $n_{\min}$ is determined by the equation $y_{n_{\min}-1} < y
< y_{n_{\min}}$.}:
  \be
d\sigma=\sum\limits_{n}d\sigma^{(n)}\,.
 \label{18}
 \ee

\subsection{Invariant polarization parameters}

The invariant description of the polarization properties of both
the initial and the final photons can be performed in the standard
way (see \cite{BLP}, \S 87). We define a pair of unit 4-vectors
\begin{equation}
e^{(1)}={N\over\sqrt{-N^2}}\,,\;\;\;
e^{(2)}={P\over\sqrt{-P^2}}\,,
 \label{19}
\end{equation}
where\footnote{Below we use the system of units in which $c=1$,
$\hbar=1$.}
 \bea
N^\mu&=&\varepsilon^{\mu\alpha \beta \gamma} P_\alpha
(k'-n\,k)_\beta K_\gamma \, ,
 \nn
 \\
 P_\alpha&=&(q+q')_\alpha-{(q+q')K\over
K^2}\,K_\alpha \,,\;\;\; K_\alpha=n\,k_\alpha+k'_\alpha\,,
 \nn\\
\sqrt{-N^2}&=&m^3xy\, {n\,s_n\over r_n} \sqrt{1+\xi^2}\,,\;\;\;
\sqrt{-P^2}=m \, {s_n\over r_n} \sqrt{1+\xi^2}\,.
 \nn
 \eea
The 4-vectors $e^{(1)}$ and $e^{(2)}$ are orthogonal to each other
and to the 4-vectors $k$ and $k'$,
 $$
e^{(i)} e^{(j)} = - \delta_{ij}, \;\; e^{(i)} k = e^{(i)} k' =0;
\;\; i,\, j = 1,\,2 \,.
 $$
Therefore, they are fixed with respect to the scattering plane of
the process.

Let $\xi_j$ be the Stokes parameters for the initial photon which
are defined with respect to 4-vectors $e^{(1)}$ and $e^{(2)}$. As
for the polarization of the final photon, it is necessary to
distinguish the polarization $\xi_j^{(f)}$ of the final photon as
resulting from the scattering process itself from the detected
polarization $\xi '_j$ which enters the effective cross section
and which essentially represents the properties of the detector as
selecting one or other polarization of the final photon (for
detail  see~\cite{BLP}, \S 65). Both these Stokes parameters,
$\xi_j^{(f)}$ and $\xi '_j$, are also defined with respect to the
4-vectors $e^{(1)}$ and $e^{(2)}$.

Let {\boldmath $\zeta$} be the polarization vector of the initial
electrons. As with the final photon, it is necessary to
distinguish the polarization {\boldmath $\zeta$}$^{(f)}$ of the
final electron as such from the polarization {\boldmath $\zeta
$}$^{\prime}$ that is selected by the detector. The vectors
{\boldmath $\zeta$} and {\boldmath $\zeta $}$^{\prime}$ enter the
effective cross section. They also determine the electron-spin
4-vectors
 \bea
a&=&\left({\mbox {\boldmath $\zeta$}{\bf p}\over m}\,,\,
\mbox{\boldmath $\zeta$}+ {\bf p}{\mbox{\boldmath $\zeta$} {\bf p}
\over m(E+m)}\right) \,,
 \nn
 \\
a^{\prime}&=&\left({\mbox {\boldmath$\zeta$}^{\prime} {\bf
p}^{\prime}\over m}\,,\, \mbox {\boldmath$\zeta$}^{\prime} +{\bf
p}^{\prime}{\mbox {\boldmath$\zeta$}^{\prime} {\bf
p}^{\prime}\over m(E^{\prime}+m)}\right)
 \label{21}
 \eea
and the mean helicity of the initial and final electrons
\begin{equation}
\lambda_e={\mbox{\boldmath $\zeta$}{\bf p}\over 2|{\bf
p}|}\,,\;\;\;\; \lambda^{\prime}_e={\mbox {\boldmath$
\zeta$}^{\prime} {\bf p}^{\prime}\over 2|{\bf p}^{\prime}|}\,.
 \label{22}
\end{equation}

Now we have to define invariants which describe the polarization
properties of the initial and the final electrons. For the
electrons it is a more complicated task than it is for the
photons. For the linear Compton scattering, the relatively simple
description was obtained in ~\cite{Grozin} using invariants which
have a simple meaning in the center-of-mass system. However, this
frame of reference is not convenient for the description of the
non-linear Compton scattering, since it has actually to vary with
the change of the number of the absorbed laser photons $n$.

Since $ap=a'p'=0$, we can decompose the electron polarization
4-vectors over three convenient unit 4-vectors $e_j$ and $e_j'$,
$j=1,\,2,\,3$, projections on which determine polarization
properties of the electrons. Our choice is based on the experience
obtained in~\cite{KPS98} and \cite{BPM}. We exploit two ideas.
First of all, the 4-vector $e^{(1)}$ is orthogonal to the
4-vectors $k,\;k',\; p$ and $p'$, therefore, the invariants
$\zeta_1 =-ae^{(1)}$ and $\zeta'_1 =-a'e^{(1)}$ are the transverse
polarizations of the initial and the final electrons perpendicular
to the scattering plane. Further, it is not difficult to check
that the invariant $ak/(mx)$ is the mean helicity of the initial
electron in the frame of reference, in which the electron momentum
${\bf p}$ is anti-parallel to the initial photon momentum ${\bf
k}$. Analogously, the invariant $a'k/[(mx(1-y)]$ is the mean
helicity of the final electron in the frame of reference, in which
the momentum of the final electron ${\bf p}'$ is anti-parallel to
the initial photon momentum ${\bf k}$. It is important that this
interpretation is valid for any number of absorbed laser photons.
Furthermore, we will show that for the practically important case,
relevant for the $e \to \gamma$ conversion, these frames of
reference almost coincide.

As a result, we define the two sets of units 4-vectors
 \bea
e_1&=&e^{(1)}\,,\;\;\; e_2=-e^{(2)}-{\sqrt{-P^2}\over m^2 x}k\,,
 \nn
 \\
e_3&=&{1\over m}\left(p-{2\over x}k\right);
 \label{23}\\
e^{\prime}_1&=&e^{(1)}\,,\;\;\;e^{\prime}_2=-e^{(2)}-{\sqrt{-P^2}
\over m^2 x(1-y)}k\,,
 \nn
 \\
e^{\prime}_3&=&{1\over m}\left(p^{\prime}-{2\over x(1-y)}k\right).
 \nn
 \eea
These vectors satisfy the conditions
 \begin{equation}
e_i e_j=-\delta_{ij}\;,\;\;\;e_j p=0\; ; \;\;\;e^{\prime}_i
e^{\prime}_j= -\delta_{ij}\; , \;\;\; e^{\prime}_j p^{\prime}=0\,.
 \label{24}
\end{equation}

It allows us to represent the 4-vectors $a$ and $a^{\prime}$ in
the following covariant form:
\begin{equation}
a=\sum_{j=1}^3 \zeta_j e_j\,,\;\;\; a^{\prime}=\sum_{j=1}^3
\zeta^{\prime}_j e^{\prime}_j \,,
 \label{25}
\end{equation}
where
 \be
 \zeta_j=-ae_j\,,\;\;\;
\zeta^{\prime}_j=-a^{\prime}e^{\prime}_j\,.
 \label{26}
\end{equation}
The invariants $\zeta_j$ and $\zeta^{\prime}_j$ describe
completely the polarization properties of the initial electron and
the detected polarization properties of the final electron,
respectively. To clarify the meaning of invariants $\zeta_j$, it
is useful to note that
 \begin{equation}
\zeta_j ={\mbox{\boldmath $\zeta$}}{\bf n}_j\,,
 \label{27}
\end{equation}
where the corresponding 3-vectors are
\begin{equation}
{\bf n}_j ={\bf e}_j -{{\bf p}\over E+m}\;e_{j0}
 \label{28}
\end{equation}
with $e_{j0}$ being a time component of the 4-vector $e_{j}$
defined in (\ref{23}).  Using the properties (\ref{24}) of the
 4-vectors $e_j$,
one can check that
\begin{equation}
{\bf n}_i \; {\bf n}_j =\delta_{ij}\,.
 \label{29}
\end{equation}
As a result, the polarization vector  $\mbox{\boldmath$\zeta$}$
has the form
\begin{equation}
\mbox{\boldmath$\zeta$}=\sum^3_{j=1} \zeta_j\;{\bf n}_j\, .
 \label{30}
\end{equation}
Analogously, for the final electron the invariants
$\zeta^{\prime}_j$ can be presented as
\begin{equation}
\zeta'_j ={\mbox{\boldmath $\zeta$}}'{\bf n}'_j, \;\;\; {\bf n}'_j
={\bf e}'_j -{{\bf p}'\over E'+m}\;e'_{j0}, \;\;\; {\bf n}'_i \;
{\bf n}'_j =\delta_{ij}\,,
 \label{31}
\end{equation}
such that
\begin{equation}
\mbox{\boldmath$\zeta$}'=\sum^3_{j=1} \zeta'_j\;{\bf n}'_j\, .
 \label{32}
\end{equation}

In what follows, we will often consider the non-linear Compton
scattering in the frame of reference in which an electron performs
a head-on collision with laser photons, i.e. in which ${\bf p}
\,\parallel \,(- {\bf k})$. We call this the ``collider system''.
In this frame of reference we choose the $z$-axis along the
initial electron momentum ${\bf p}$. Azimuthal angles $\varphi,\;
\beta$ and $\beta'$ of vectors ${\bf k}',\;
\mbox{\boldmath$\zeta$}$ and $\mbox{\boldmath$\zeta$}'$ are
defined with respect to one fixed $x$-axis.

\section{Cross section in the invariant form}

\subsection{General relations}

The effective differential cross section can be presented in the
following invariant form:
 \bea
&&d\sigma (\mbox{\boldmath$\zeta$},\,\mbox{\boldmath$\xi$},\,
\mbox{\boldmath$\zeta$}^{\prime},\,\mbox{\boldmath$\xi$}^{\prime})
= {r_e^2\over 4x}\;\sum_n F^{(n)}\;d\Gamma_n\,,
 \label{33}
 \\
&&d\Gamma_n = \delta (q+n\,k-q'-k')\;{d^3k'\over
\omega'}{d^3q'\over q_0'}\,,
 \nn
 \eea
where $r_e=\alpha/m$ is the classical electron radius, and
 \bea
F^{(n)}&=&F_0^{(n)}+\sum ^3_{j=1}\left( F_j^{(n)}\xi '_j\; +
\;G_j^{(n)} \zeta^{\prime}_j\right)
 \nn
 \\
&+& \sum ^3_{i,j=1}H_{ij}^{(n)}\,\zeta^{\prime}_i\,\xi^{\prime}_j
\,.
 \label{34}
 \eea
Here the function $F_0^{(n)}$ describes the total cross section
for a given harmonic $n$, summed over spin states of the final
particles:
 \be
\sigma^{(n)}(\mbox{\boldmath$\zeta$},\,\mbox{\boldmath$\xi$})=
{r_e^2\over x}\; \int F_0^{(n)}\,d\Gamma_n \,.
 \label{35}
 \ee
The terms $F_j^{(n)}\xi '_j$  and $G_j^{(n)} \zeta^{\prime}_j$ in
(\ref{34}) describe the polarization of the final photons and the
final electrons, respectively. The last terms $H_{ij}^{(n)}
\zeta^{\prime}_i\,\xi^{\prime}_j$ stand for the correlation of the
final particles' polarizations.

From (\ref{33}), (\ref{34}) one can deduce the polarization of the
final photon $\xi_j^{(f)}$ and electron $\zeta^{(f)}_j$ resulting
from the scattering process itself. According to the usual rules
(see~\cite{BLP}, \S 65), we obtain the following expression for
the Stokes parameters of the final photon (summed over
polarization states of the final electron):
 \bea
\xi_j^{(f)}&=& {F_j\over F_0}\,,\;\; F_0=\sum_n
F_0^{(n)}\,,\;\;F_j=\sum_n F_j^{(n)}\,;
 \nn
 \\
j&=& 1,\,2,\,3\,.
 \label{36}
 \eea
The polarization of the final electron
(summed over polarization states of the final photon)
is given by invariants
 \be
\zeta_j^{(f)}= {G_j\over F_0}\,,\;\;G_j=\sum_n G_j^{(n)}\,,
 \label{37}
 \ee
therefore, its polarization vector is
 \be
\mbox{ \boldmath $\zeta$}^{(f)}= \sum_{j=1}^3 \, {G_j\over F_0}\,
{\bf n}_j^{\prime}\,.
 \label{38}
 \ee
In the similar way, the polarization properties for a given
harmonic $n$ are described by
\begin{equation}
\xi_{j}^{(n)(f)}= {F_j^{(n)}\over F_0^{(n)}}\,,\;\;
\zeta_{j}^{(n)(f)}= {G_j^{(n)}\over F_0^{(n)}}\,.
 \label{39}
\end{equation}

\subsection{The results for the circularly polarized laser photons}

In this subsection we consider the case of 100\% circularly
polarized laser beam. The electromagnetic laser field is described
by the 4-potential
 \be
A_\mu(x) ={m\over e}\,\xi \;\left[e_\mu^{(1)}\,
\cos{(kx)}+P_c\,e_\mu^{(2)}\, \sin{(kx)} \right]\,,
 \label{2.2a}
 \ee
where the unit vectors $e_\mu^{(1,2)}$ are given in (\ref{19}) and
$P_c$ is the degree of the circular polarization of the laser wave
or the initial photon helicity. Therefore, the Stokes parameters
of the laser photon are
 \be
\xi_1=\xi_3=0,\;\; \xi_2=P_c= \pm 1\,.
 \label{20}
 \ee

We have calculated the coefficients $F_j^{(n)},\, G_j^{(n)}$ and
$H_{ij}^{(n)}$ using the standard technique presented
in~\cite{BLP}, \S 101. The necessary traces have been calculated
using the package MATHE\-MA\-TICA. In the considered case of the
100 \% circularly polarized ($P_c=\pm 1$) laser beam, almost all
dependence on the non-linearity parameter $\xi^2$ accumulates in
three functions:
  \bea
f_n& \equiv&
f_n(z_n)=J_{n-1}^{2}(z_n)+J_{n+1}^{2}(z_n)-2J_{n}^{2}(z_n)\,,
 \nn\\
 g_n& \equiv& g_n(z_n)=\frac{4 n^2 J_{n}^2(z_n)}{z_n^2}\,,
 \label{40}\\
h_n&\equiv& h_n(z_n)=J_{n-1}^{2}(z_n)-J_{n+1}^{2}(z_n)\,,
 \nn
 \eea
where $J_n(z)$ is the Bessel function. The functions (\ref{40})
depend on $x$, $y$ and $\xi$ via the single argument
 \be
z_n= {\xi\over \sqrt{1+\xi^2}}\; n\,s_n\,.
 \label{41}
 \ee
For the small value of this argument one has
\be
f_n=g_n=h_n=\frac{(z_n/2)^{2(n-1)}}{[(n-1)!]^2} \;\; \mbox{ at}
\;\; z_n \to 0\,,
 \label{42}
 \ee
in particular,
 \be
 f_1=g_1=h_1=1
\;\; \mbox{ at} \;\; z_1= 0\,.
 \label{43}
 \ee
It is useful to note that this argument is small for small
$\xi^2$, as well as for small or for large values of $y$:
 \bea
 z_n &\to& 0 \;\; \mbox{either at} \;\; \xi^2\to 0\,,
 \nn
 \\
 &&\mbox{ or at }\;\; y\to 0\,,\;\;\mbox{ or at }\;\; y\to y_n\,.
  \label{44}
 \eea

The results of our calculations are the following. The function
$F_0^{(n)}$, related to the total cross section (\ref{35}), reads
 \bea
F_0^{(n)}&=&\left({1\over 1-y}+1-y\right)\,f_n- {s_n^2\over
1+\xi^2}\, g_n
 \nn
 \\
&-& \left[{y s_n \over \sqrt{1+\xi^2}}\, \zeta_2 -{y(2-y)\over
1-y}\, c_n \,\zeta_3\right] \,h_n\,P_c\,.
 \label{45}
 \eea

The polarization of the final photons $\xi_j^{(f)}$ is given by
Eq. (\ref{36}) where
  \bea
F_1^{(n)}&=&{y\over 1-y} {s_n \over \sqrt{1+\xi^2}}\, h_n P_c\,
\zeta_1\,,
 \label{46}
 \\
F_2^{(n)}&=&\left({1\over 1-y} +1-y\right)\, c_n h_n P_c- {ys_n
c_n \over \sqrt{1+\xi^2}}\, g_n \, \zeta_2
 \nn
 \\
&+&y\left( {2-y\over 1-y}\,f_n- {s^2_n\over 1+\xi^2}\,g_n\right)
\zeta_3\,,
 \nn
 \\
F_3^{(n)}&=&2(f_n-g_n)+ s^2_n (1+\Delta)\,g_n
 \nn
 \\
&-&{y\over 1-y} {s_n \over \sqrt{1+\xi^2}}\, h_n P_c\, \zeta_2 \,;
 \nn
 \eea
here and below we use the notation
 \be
 \Delta= {\xi^2 \over 1+\xi^2}\,.
 \label{delta}
 \ee

The polarization of the final electrons $\zeta_j^{(f)}$ is given
by (\ref{37}), (\ref{38}) with
 \bea
G_1^{(n)}&=&\, G^{(n)}_\perp\, \zeta_1,
 \label{47}\\
G_2^{(n)}&=&\, G^{(n)}_\perp\, \,\zeta_2 -
\frac{y{s}_n}{(1-y)\sqrt{1+\xi^2}}\,\left(c_n\,g_n\, \zeta_3+
h_n\, P_c\right)\,,
 \nn\\
G_3^{(n)}&=&\, G^{(n)}_\parallel+\, \frac{y {s}_n
{c}_n}{\sqrt{1+\xi^2}}\,g_n \,\zeta_2,
 \nn
  \eea
where we introduced the notation
 \bea
G_\perp^{(n)}&=& 2f_n-\frac{{s}_n^2}{1+\xi^2}\,g_n\,,
 \label{48}
 \\
G^{(n)}_\parallel &=& \left[\left(\frac{1}{1-y}
 + 1-y\right)f_n \right.
 \label{48a}
 \\
&-& \left.
\left(1+\frac{y^2}{1-y}\right)\frac{{s_n^2}}{1+\xi^2}g_n \right]
\zeta_3 + \frac{y(2-y){c}_n}{1-y}\,h_nP_c\,.
 \nn
 \eea

Finally, the correlations of the final particles' polarizations
are
 \bea
H_{11}^{(n)}&=&\frac{ys_n}{\sqrt{1+\xi^2}}h_n P_c +\frac{y}{1-y}
\left[(2-y)(f_n-g_n)\right.
 \nn
 \\
&+&\left. (1-y+\Delta)s_n^2g_n\right]\zeta_2 -\frac{yc_n
s_n}{\sqrt{1+\xi^2}}g_n \zeta_3,
 \nonumber \\
H_{21}^{(n)}&=&-\frac{y}{1-y} \left\{(2-y)(f_n-g_n) \right.
 \nn
 \\
&+&\left. \left[1+(1-y)\Delta\right] s_n^2g_n\right\}\zeta_1 \, ,
 \nonumber \\
H_{31}^{(n)}&=& \frac{y c_n s_n}{(1-y)\sqrt{1+\xi^2}}g_n \zeta_1
\, , \ \ \ H_{12}^{(n)}= 2 c_n h_n P_c\, \zeta_1 \, ,
 \nn
  \\
H_{22}^{(n)}&=&-\frac{y c_n s_n}{(1-y)\sqrt{1+\xi^2}}\,g_n+ 2 c_n
h_n P_c\, \zeta_2
 \nn
 \\
&-&\frac{y s_n}{(1-y)\sqrt{1+\xi^2}}\,h_n P_c\, \zeta_3 \, ,
 \label{49}
  \\
H_{32}^{(n)}&=&\frac{y}{1-y}
\left[(2-y)f_n-\frac{s_n^2}{1+\xi^2}g_n\right]+\frac{y s_n
}{\sqrt{1+\xi^2}}\,h_n P_c\, \zeta_2
 \nn
 \\
&+&\frac{2-2y+y^2}{1-y}\,c_n h_n P_c\, \zeta_3 \, ,
 \nonumber \\
H_{13}^{(n)}&=&\left[\frac{2-2y+y^2}{1-y}(f_n-g_n)\right.
 \nn
 \\
&+& \left.
\left(1+\frac{1-y+y^2}{1-y}\Delta\right)s_n^2\,g_n\right]\,\zeta_1
\,,
 \nonumber \\
H_{23}^{(n)}&=&-\frac{y s_n}{\sqrt{1+\xi^2}}\,h_n
P_c+\left[\frac{2-2y+y^2}{1-y}(f_n-g_n)
\right.
 \nn
 \\
&+& \left. \left(\frac{1-y+y^2}{1-y} + \Delta
\right)\,s_n^2\,g_n\right]\,\zeta_2
 \nn\\
&+& \frac{y c_n s_n}{\sqrt{1+\xi^2}}\,g_n \,\zeta_3 \, ,
 \nonumber \\
H_{33}^{(n)}&=&-\frac{y c_n s_n}{(1-y)\sqrt{1+\xi^2}}\,g_n\,
\zeta_2
 \nn
 \\
&+& \left[2(f_n-g_n)+(1+\Delta)\,s_n^2\,g_n\right]\, \zeta_3 \, .
 \nn
 \eea

\subsection{The results for the linearly polarized laser photons}

Here we consider the case of 100\% linearly polarized laser beam.
The electromagnetic laser field is described by the 4-potential
 \bea
A^\mu(x)& =& A^\mu\,\cos{(kx)}\,,\;\; A^\mu= {\sqrt{2}\,m\over
e}\,\xi \;e_{\rm L}^\mu\,,
 \nn
 \\
e_{\rm L} e_{\rm L} &=& -1\,,
 \label{2.4}
 \eea
where $e^\mu_{\rm L}$ is the unit 4-vector describing the
polarization of the laser photons, which can be expressed in the
covariant form via the unit 4-vectors $e^{(1,2)}$ given in
(\ref{19}) as follows:
 \be
e_{\rm L} = e^{(1)}\, \sin{\varphi} -e^{(2)}\, \cos{\varphi}\,.
 \label{2.5}
 \ee
The invariants
 \be
\sin{\varphi}= -e_{\rm L}\,e^{(1)}\,, \;\;\;\cos{\varphi}=e_{\rm
L}\,e^{(2)}
 \label{2.5a}
 \ee
have a simple notion in the collider system. For the problem
discussed it is convenient to choose the $x$-axis of this frame of
reference along the direction of the laser linear polarization,
i.e. along the vector ${\bf e}_{{\rm L}\perp}$. With such a
choice, the quantity $\varphi$ is the azimuthal angle of the final
photon in the collider system (analogously, the azimuthal angles
$\beta$ and $\beta'$ of the electron polarizations
$\mbox{\boldmath$\zeta$}$ and $\mbox{\boldmath$\zeta$}'$ are also
defined with respect to this $x$-axis).

The Stokes parameters $\xi_i$ are defined with respect to the
$x^{\prime}y^{\prime}z^{\prime}$-axes which are fixed to the
scattering plane. The $x'$-axis is  perpendicular to the
scattering plane:
\begin{equation}
x' \;\parallel\; {\bf k} \times {\bf k}';
 \label{2.6}
\end{equation}
the $y'$-axis is in that plane
\begin{equation}
y' \;\parallel\;{\bf k} \times ({\bf k} \times {\bf
k}')=-\omega^2{\bf k}'_{\perp}\,.
 \label{2.7}
 \end{equation}
The azimuthal angle of the linear polarization of the laser photon
in the collider system equals zero with respect to the $xyz$-axes,
and it is $\varphi+(3\pi/2)$ with respect to the $x'y'z'$-axes.
Therefore, in the considered case of the 100 \% linearly polarized
laser beam one has
\begin{equation}
\xi_1=-\sin{2\varphi},\;\;\; \xi_2=0,\;\;\;
\xi_3=-\cos{2\varphi}\,.
 \label{2.9}
\end{equation}

When calculating the effective cross section (\ref{33}), we found
that almost all dependence on the non-linearity parameter $\xi^2$
accumulates in three functions:
 \bea
\tilde{f}_n&=& 4\left[A_1(n,\,a,\,b)\right]^2 -4A_0(n,\,a,\,b)
A_2(n,\,a,\,b) \,,
 \nn\\
\tilde{g}_n &=& {4n^2\over z_n^2}\, \left[A_0(n,\,a,\,b)
\right]^2\,,
 \label{2.13}
 \\
 \tilde{h}_n &=& {4n\over a}\,A_0(n,\,a,\,b)\,A_1(n,\,a,\,b)\,,
 \nn
 \eea
where the functions $A_k(n,\,a,\,b)$ were introduced
in~\cite{NR-review} as follows
 \bea
&&A_k(n,\,a,\,b)
 \label{2.14}
 \\
&& =\int\limits_{-\pi}^{\pi} \,\cos^k{\psi}\,\exp {\left[\,{\rm
i}\left(n\psi-a \sin{\psi} + b\sin{2\psi}\right) \right]}\,
{d\psi\over 2\pi}\,.
 \nn
 \eea
The arguments of these functions are
 \be
a= e\, \left( {Ap\over kp}- {Ap{^\prime}\over kp'}\right)\,,\;\;
b= {1\over 8}\, e^2 A^2 \,\left( {1\over kp}- {1\over kp'}\right)
 \label{2.15}
  \ee
or, expressed in terms of the non-linearity parameter, they are
 \be
a= \sqrt{2}\,\xi\,m\, \left( {e_{\rm L} p\over kp}- {e_{\rm L}
p'\over kp'}\right)\,,\;\; b=  {y\over 2(1-y)x}\,\xi^2\,.
 \label{2.16}
 \ee
In the collider system one has
 \be
 a=- z_n \; \sqrt{2}\, \cos{\varphi}\,,
 \label{2.17}
 \ee
where $z_n$ is defined in (\ref{41}). From the definition of the
functions (\ref{2.14}) one can deduce that
 $$
A_k(n,\,-a,\,b)=(-1)^{n+k}\,A_k(n,\,a,\,b)\,;
 $$
therefore, the functions $\tilde{f}_n$, $\tilde{g}_n$ and
$\tilde{h}_n$ are even functions of the variable $\cos{\varphi}$.
To find the photon spectrum, one needs also the functions
(\ref{2.13}) averaged over the azimuthal angle $\varphi$:
 \be
\langle \tilde{f}_n \rangle =  \int_0^{2\pi} \tilde{f}_n\, {d
\varphi \over 2\pi}\,,\;\; \langle \tilde{g}_n \rangle =
\int_0^{2\pi} \tilde{g}_n\, {d \varphi \over 2\pi}\,.
 \label{2.19}
 \ee

Among the functions $A_0$, $A_1$ and $A_2$ there are useful
relations (see~\cite{NR-review})
 \bea
A_1(n,\,a,\,b)&=&{1\over 2}
\left[A_0(n-1,\,a,\,b)+A_0(n+1,\,a,\,b)\right]\,,
 \nn
 \\
A_2(n,\,a,\,b)&=&{1\over 4}
\left[A_0(n-2,\,a,\,b)+2A_0(n,\,a,\,b)\right.
  \label{2.18}
 \\
&+&\left. A_0(n+2,\,a,\,b)\right] \,,
 \nn
  \eea
  $$
(n-2b)\,A_0(n,\,a,\,b)-a\, A_1(n,\,a,\,b)+4b \,A_2(n,\,a,\,b)=0\,.
 $$

To find the behavior of $A_0(n,\,a,\,b)$ for small values of its
arguments, one can use the expansion into the series of Bessel
functions
 \be
A_0(n,\,a,\,b)=\sum_{s=-\infty}^{s=+\infty}\,
J_{n+2s}(a)\,J_s(b)\,.
 \ee
As a consequence, for small values of $\xi^2 \to 0$ or $y \to 0$
we have
 \be
a\propto \sqrt{y\,\xi^2}\,,\;\;b \propto {y\,\xi^2}\,, \;\;
A_{k}(n,\,a,\,b) \propto \left( {y\,\xi^2}\, \right)^{|n-k|/2}
  \label{2.20a}
  \ee
  and
 \be
\tilde{f}_n ,\,\tilde{g}_n ,\, \tilde{h}_n \propto
\left({y\,\xi^2}\right)^{n-1} \, .
 \label{2.20}
 \ee
In particular, at $\xi^2=0$ or at  $y=0$
 \be
\tilde{f}_1=\langle \tilde{f}_1 \rangle= \langle \tilde{g}_1
\rangle= \tilde{h}_1=1\,,\;\; \tilde{g}_1=1+\cos2\varphi\,.
 \label{2.21}
 \ee

The results of our calculations are the following. First, we
define the auxiliary functions
 \bea
X_n&=& \tilde{f}_n-(1+c_n)\left[ (1-\Delta\, r_n)\,\tilde{g}_n
-\tilde{h}_n\,\cos{2\varphi}\right] \, ,
 \nn \\
Y_n&=&(1+c_n)\tilde{g}_n-2\tilde{h}_n \cos^2\!{\varphi} \, ,
 \label{2.22}
\\
V_n&=& \tilde{f}_n\,\cos{2\varphi}
 \nn
 \\
 &+& 2 (1+c_n) \left[(1-\Delta\,
r_n)\,\tilde{g}_n - 2\tilde{h}_n\, \cos^2\!{\varphi}\,
\right]\,\sin^2\!{\varphi} \,,
 \nn
 \eea
where $c_n$, $r_n$ and $\Delta$ are defined in (\ref{10}),
(\ref{11}) and (\ref{delta}), respectively.

The term $F_0^{(n)}$, related to the total cross section
(\ref{35}), reads
 \be
F_0^{(n)}=\left({1\over 1-y}+1-y\right)\,\tilde{f}_n- {s_n^2\over
1+\xi^2}\, \tilde{g}_n \ .
 \label{2.23}
 \ee

 The polarization of the final photons $\xi_j^{(f)}$ is
given by (\ref{36}), where
  \bea
F_1^{(n)}&=&2\, X_n\, \sin{2\varphi} \,,\;\;\; F_3^{(n)}= -2\,
V_n+\frac{s_n^2}{1+\xi^2}\,\tilde{g}_n \,,
 \nn
 \\
F_2^{(n)}&=& {ys_n \over \sqrt{1+\xi^2}} \left( \, \tilde{h}_n \,
\zeta_1\, \sin{2\varphi}-Y_n\, \zeta_2 \right)
 \label{2.24}
 \\
 &+&y\left( {2-y\over
1-y}\,\tilde{f}_n- {s^2_n\over 1+\xi^2}\,\tilde{g}_n\right) \,
\zeta_3\,,
 \nn
 \eea

The polarization of the final electrons $\zeta_j^{(f)}$ is given
by (\ref{37}), (\ref{38}) with
 \bea
G_1^{(n)}&=& \tilde{G}^{(n)}_\perp\,\zeta_1+ {ys_n \over (1-y)
\sqrt{1+\xi^2}}\, \tilde{h}_n \, \zeta_3 \,\sin{2\varphi}\, ,
 \nn\\
G_2^{(n)}&=&\tilde{G}^{(n)}_\perp\,\zeta_2- {ys_n \over (1-y)
\sqrt{1+\xi^2}}\, Y_n\zeta_3\,,
 \label{2.25}\\
G_3^{(n)}&=& \tilde{G}^{(n)}_\parallel - {ys_n \over
\sqrt{1+\xi^2}} \left( \, \tilde{h}_n \, \zeta_1\, \sin{2\varphi}-
Y_n \zeta_2 \right)\,,
 \nn
 \eea
where we use the notation
 \bea
\tilde{G}^{(n)}_\perp&=&2\tilde{f}_n-\frac{{s}_n^2}{1+\xi^2}\,\tilde{g}_n
 \label{2.25a}
 \\
\tilde{G}^{(n)}_\parallel &=&\left[\left(\frac{1}{1-y}+1-y\right)
\tilde{f}_n \right.
 \nn
 \\
 &-& \left. \left(1+\frac{y^2}{1-y}\right)
\frac{{s_n^2}}{1+\xi^2}\tilde{g}_n \right]\, \zeta_3\, .
 \label{2.25ab}
  \eea

At last, the correlations of the final particles' polarizations
are
 \bea
 H_{11}^{(n)}&=&\frac{2-2y+y^2}{1-y}\, X_n \zeta_1\,\sin{2\varphi}
 \nn
 \\
&-& y\! \left( \frac{2-y}{1-y}\, V_n - {s_n^2
 \over 1+\xi^2}\, \tilde{g}_n \right) \zeta_2 -{ys_n
 \over \sqrt{1+\xi^2}}Y_n   \zeta_3 \, ,
  \nonumber \\
H_{21}^{(n)}&=& \frac{y}{1-y} \left[ (2-y)V_n
-\frac{s_n^2}{1+\xi^2}\, \tilde{g}_n \right]\zeta_1
 \nn
 \\
&+&\frac{2-2y+y^2}{1-y} X_n \zeta_2\,\sin{2\varphi}
  \nn \\
&& + {ys_n \over \sqrt{1+\xi^2}} \tilde{h}_n \, \zeta_3
\,\sin{2\varphi}\, ,
 \nonumber \\
H_{31}^{(n)}&=&{ys_n  \over (1-y)\sqrt{1+\xi^2}}\left( Y_n \zeta_1
- \tilde{h}_n \,  \zeta_2\, \sin{2\varphi}\right)
 \nn
 \\
 &+&2 X_n \zeta_3
\,\sin{2\varphi}\, ,
  \nn \\
H_{12}^{(n)}&=& \frac{y s_n}{(1-y)\sqrt{1+\xi^2}}\, \tilde{h}_n \,
\sin{2\varphi}\,,
 \nn
 \\
H_{22}^{(n)}&=& -\frac{y s_n}{(1-y)\sqrt{1+\xi^2}}\, Y_n \, ,
 \nn
  \\
H_{32}^{(n)}&=&\frac{y}{1-y}\left[
(2-y)\tilde{f}_n-\frac{s_n^2}{1+\xi^2}\, \tilde{g}_n \right] \, ,
  \label{2.26} \\
H_{13}^{(n)}&=& -
\left(\frac{2-2y+y^2}{1-y}V_n-\frac{s_n^2}{1+\xi^2}\,
\tilde{g}_n\right)\zeta_1
 \nn
 \\
&-&\frac{y(2-y)}{1-y}\,X_n\zeta_2\,\sin{2\varphi}
   \nn \\
&& + \frac{y s_n}{\sqrt{1+\xi^2}}\, \tilde{h}_n\,
\zeta_3\,\sin{2\varphi}\, ,
  \nn \\
H_{23}^{(n)}&=& \frac{y(2-y)}{1-y}\,X_n\zeta_1\,\sin{2\varphi}
-\left(\frac{2-2y+y^2}{1-y}V_n \right.
 \nn
 \\
&-& \left. \frac{1-y+y^2}{1-y}\frac{s_n^2}{1+\xi^2} \,
\tilde{g}_n\right)\zeta_2
  \nn\\
&& +\frac{y s_n}{\sqrt{1+\xi^2}}Y_n \zeta_3\, ,
  \nn \\
H_{33}^{(n)}&=&-\frac{y s_n}{(1-y)\sqrt{1+\xi^2}} \left( \,
\tilde{h}_n\, \zeta_1\,\sin{2\varphi} + Y_n \zeta_2 \right)
 \nn
 \\
&-&\left( 2\, V_n - \frac{s_n^2}{1+\xi^2}\,
\tilde{g}_n\right)\zeta_3 \, .
  \nn
  \eea

\section{Going to the collider system}

\subsection{Exact relations}

As an example of the application of the above formulae, let us
consider the non-linear Compton scattering in the collider system
defined in Sect. 2.3. In such a frame of reference the unit
vectors ${\bf n}_j$, defined in (\ref{28}), has a simple form:
 \bea
{\bf n}_1&=&{{\bf p}\times {\bf p}'\over |{\bf p}\times {\bf
p}'|}= {{\bf k}\times {\bf k}'\over |{\bf k}\times {\bf
k}'|}\,,\;\;\; {\bf n}_2={{\bf p}\times {\bf n}_1\over |{\bf
p}\times{\bf n}_1|}= {{\bf k}'_{\perp} \over |{\bf
k}'_{\perp}|}\,,
 \nn
 \\
{\bf n}_3&=&{{\bf p}\over |{\bf p}|}\,,
 \label{50}
 \eea
and the invariants $\zeta_j$ are equal to
 \bea
\zeta_1 &=&\mbox{\boldmath$\zeta$}{\bf n}_1 =\zeta_{\perp}
\sin{(\varphi-\beta)}\,,\;\;\; \zeta_2 =
\mbox{\boldmath$\zeta$}{\bf n}_2 =\zeta_{\perp}
\cos{(\varphi-\beta)} \,,
 \nn
 \\
\zeta_3& =& \mbox{\boldmath$\zeta$}{\bf n}_3 =2\lambda_e\,.
 \label{51}
 \eea
Here ${\bf k}'_{\perp}$ and $\mbox{\boldmath$\zeta$} _{\perp}$
stands for the transverse components of the vectors ${\bf k}'$ and
$\mbox{\boldmath $\zeta$}$ with respect to the vector ${\bf p}$,
and $\zeta_{\perp} = |\mbox{\boldmath$\zeta$} _{\perp}|$.
Therefore, in this frame of reference, $\zeta_1$ is the transverse
polarization of the initial electron perpendicular to the
scattering plane, $\zeta_2$ is the transverse polarization in that
plane and $\zeta_3$ is the doubled mean helicity of the initial
electron.

A phase volume element in (\ref{33}) is equal to
\begin{equation}
d\Gamma=dy\;d\varphi\,,
 \label{52}
\end{equation}
and the differential cross section, summed over spin states of the
final particles, is
 \be
{d\sigma^{(n)}(\mbox{\boldmath$\zeta$},\,\mbox{\boldmath$\xi$})\over
dy \, d\varphi} = {r_e^2\over x}\; F_0^{(n)} \,.
 \label{53}
 \ee
Integrating this expression over $\varphi$ and then over $y$, we
find
 \bea
{d\sigma^{(n)}(\mbox{\boldmath$\zeta$},\,\mbox{\boldmath$\xi$})\over
dy}& =& {2\pi r_e^2\over x}\, \langle F_0^{(n)} \rangle \,,
  \label{54}
  \\
\sigma^{(n)}(\mbox{\boldmath$\zeta$},\,\mbox{\boldmath$\xi$})
&=&{2\pi r_e^2\over x}\, \int_0^{y_n} \langle F_0^{(n)} \rangle \,
dy\,,
 \nn
 \eea
where for the circular polarization
 \bea
\langle F_0^{(n)} \rangle&=& \left({1\over 1-y}+1-y\right)\,f_n-
{s_n^2\over 1+\xi^2}\, g_n
 \nn
 \\
&+&{y(2-y)\over 1-y}\, c_n \,h_n\,\zeta_3\,P_c
 \label{54c}
 \eea
and for the linear polarization
 \be
\langle F_0^{(n)} \rangle= \left({1\over 1-y}+1-y\right)\,\langle
\tilde{f}_n \rangle- {s_n^2\over 1+\xi^2}\,\langle \tilde{g}_n
\rangle\,.
 \label{54cl}
 \ee
This means that for the circularly polarized laser photons the
differential $d\sigma^{(n)}(\mbox{\boldmath$\zeta$},
\,\mbox{\boldmath$\xi$}) / dy$ and the total
$\sigma^{(n)}(\mbox{\boldmath$\zeta$},\,\mbox{\boldmath$\xi$})$
cross sections for a given harmonic $n$ do not depend on the
transverse polarization of the initial electron. For the linearly
polarized laser photons, these cross sections do not depend at all
on the polarization of the initial electron. These properties are
similar to those in the linear Compton scattering.

The polarization of the final photon is given by (\ref{36}). The
polarization vector of the final electron is determined by
(\ref{38}), but, unfortunately, the unit vectors ${\bf n}'_j$ in
this equation have no simple form similar to (\ref{50}).
Therefore, we have to find the characteristics that are usually
used for description of the electron polarization --- the mean
helicity of the final electron $\lambda_e'$ and its transverse (to
the vector ${\bf p}'$) polarization $\mbox{ \boldmath
$\zeta$}^{\prime}_\perp$ in the considered collider system. For
this purpose we introduce unit vectors $\mbox{ \boldmath$\nu$}_j$,
one of them is directed along the momentum of the final electron
${\bf p}^{\prime}$ and two others are in the plane transverse to
this direction (compare with (\ref{50})):
 \bea
&&\mbox{\boldmath$\nu$}_1 = {\bf n}_1, \;\;\;
\mbox{\boldmath$\nu$}_2 ={{\bf p}'\times {\bf n}_1 \over|{\bf
p}'\times{\bf n}_1|}, \;\;\; \mbox{\boldmath$\nu$}_3 = {{\bf
p}'\over |{\bf p}'|}\,;
 \nn
 \\
&&\mbox{\boldmath$\nu$}_i \, \mbox{\boldmath$\nu$}_j =
\delta_{ij}\,.
 \label{55}
 \eea
Therefore, $\mbox{\boldmath$\zeta$}'\,\mbox{\boldmath$\nu$}_1$ is
the transverse polarization of the final electron perpendicular to
the scattering plane,
$\mbox{\boldmath$\zeta$}'\,\mbox{\boldmath$\nu$}_2$ is the
transverse polarization in that plane and
$\mbox{\boldmath$\zeta$}'\,\mbox{\boldmath$\nu$}_3$ is the doubled
mean helicity of the final electron:
 \be
\mbox{\boldmath$\zeta$}'\,\mbox{\boldmath$\nu$}_3= 2\lambda'_e\,.
 \label{56}
 \ee

Let us discuss the relation between the projection
$\mbox{\boldmath$\zeta$}'\mbox{\boldmath$\nu$}_j$, defined above,
and the invariants $\zeta_j'$, defined in (\ref{26}), (\ref{31}).
In the collider system the vectors $\mbox{\boldmath$\nu$}_1$ and
${\bf n}_1'$ coincide
 \be
\mbox{\boldmath$\nu$}_1={\bf n}_1'\,,
 \ee
therefore,
 \be
\mbox{\boldmath$\zeta$}'\,\mbox{\boldmath$\nu$}_1 = \zeta_1'\,.
 \ee
Two other unit vectors ${\bf n}_2'$ and ${\bf n}_3'$ are in the
scattering plane and they can be obtained from the vectors
$\mbox{\boldmath$\nu$}_2$ and $\mbox{\boldmath$\nu$}_3$ by the
rotation around the axis $\mbox{\boldmath$\nu$}_1$ on the angle
$(-\Delta \theta)$:
 \bea
\mbox{\boldmath$\zeta$}'\,\mbox{\boldmath$\nu$}_2&=&
\zeta'_2\,\cos{\Delta \theta}+ \zeta'_3\,\sin{\Delta \theta}\,,
 \nn
 \\
\mbox{\boldmath$\zeta$}'\,\mbox{\boldmath$\nu$}_3&=&
\zeta'_3\,\cos{\Delta \theta}- \zeta'_2\,\sin{\Delta \theta}\,,
 \label{59}
 \eea
where
 \be
\cos{\Delta \theta}= \mbox{\boldmath$\nu$}_3{\bf n}_3'\,,\;\;
\sin{\Delta \theta}= - \mbox{\boldmath$\nu$}_3{\bf n}_2'\,.
 \ee
As a result, the polarization vector of the final electron
(\ref{38}) is expressed as follows:
 \bea
\mbox{\boldmath$\zeta$}^{(f)}&=&
\mbox{\boldmath$\nu$}_1\,{G_1\over F_0}\,+ \mbox{\boldmath$\nu$}_2
\left( {G_2\over F_0} \cos{\Delta \theta} + {G_3\over F_0}
\sin{\Delta \theta} \right)
 \nn
 \\
 &+&
\mbox{\boldmath$\nu$}_3 \left( {G_3\over F_0} \cos{\Delta \theta}
- {G_2\over F_0} \sin{\Delta \theta} \right)\,.
 \label{61}
 \eea

At the end of this subsection we consider the relation between the
used collider system and the centre-of-mass system (CMS) for a
given harmonic $n$, defined by the requirements:
  \be
{\bf p}'\,=- {\bf k}'\,,\;\;
 {\bf p}\,= - n\,(1-r_n\Delta)\,{\bf k}
  \label{a7b}
  \ee
(in accordance with the conservation law (\ref{8}) rewritten in
the form $p+n(1-r_n\Delta)k=p'+k'$). In this frame of reference
the invariants $\zeta'_j({\rm CMS})$ are defined analogously to
the invariants $\zeta_j$:
 \be
\zeta'_j({\rm CMS})=- a'_j e'_j({\rm CMS})\,,
 \label{23b}
 \ee
where
 \bea
e'_1({\rm CMS})&=&e^{(1)}\,,\;\;
 e^{\prime}_2({\rm CMS})=e^{(2)}+{\sqrt{-P^2}
\over 2p'k'}k'\,,
 \nn
 \\
e^{\prime}_3({\rm CMS})&=&{1\over m}\left(p^{\prime}-{m^2\over
p'k'}k'\right)\,.
 \eea
It is not difficult to find the relation between these invariants
and the invariants $\zeta'_j$, defined in (\ref{26}):
 \bea
\zeta_1'({\rm CMS})&=& \zeta_1'\,,\;\; \zeta_2'({\rm CMS})=
\cos{\theta^*_n} \,\zeta_2' +\sin{\theta^*_n} \,\zeta_3'\,,
 \nn
 \\
\zeta_3'({\rm CMS})&=& \cos{\theta^*_n} \,\zeta_3'
-\sin{\theta^*_n }\,\zeta_2' \,,
 \label{a8b}
 \eea
where $\theta^*_n$ is the photon scattering angle in the rest
system of the final electron (${\bf p}'=0$)
 \bea
\cos{\theta^*_n}&=&1-{2r_n\over 1+(1-r_n)\,\xi^2}\,,
 \nn
 \\
\sin{\theta^*_n} &=&{s_n\over (1-r_n \Delta)\,\sqrt{1+\xi^2} }\,,
 \label{a9b}
 \eea
with $\Delta$ defined in (\ref{delta}).

\subsection{Approximate formulae}

All the above formulae are exact. In  this subsection we give some
approximate formulae useful for application to the important case
of high-energy $\gamma \gamma$ and $\gamma e$ colliders. It is
expected (see, for example, the TESLA project~\cite{TESLA}) that
in the conversion region of these colliders, an electron with the
energy $E\sim 100$ GeV performs a head--on collisions with laser
photons having the energy $\omega \sim 1$ eV per a single photon.
In this case the most important kinematic range corresponds to
almost back-scattered final photons, i. e. the initial electron is
ultra-relativistic and the final photon is emitted at a small
angle $\theta_\gamma$ with respect to the $z$-axis (chosen along
the momentum ${\bf p}$ of the initial electron):
\begin{equation}
E \gg m, \;\;\; \theta_\gamma \ll 1.
 \label{62}
\end{equation}
In this approximation we have
\begin{equation}
x\approx {4E\omega\over m^2}\,,\;\;\;\; y\approx{\omega'\over E}
\approx 1- {E'\over E}\,,
 \label{63}
\end{equation}
therefore, (\ref{53}) gives us the distribution of the final
photons over the energy and the azimuthal angle. Besides, the
photon emission angle for the reaction (\ref{2}) is (see Fig.
\ref{f1})
 \be
 \theta_\gamma \approx {m\over E}\, \sqrt{nx+1+\xi^2}\;\sqrt{{y_n\over
 y}-1}
 \label{64}
 \ee
and $\theta_\gamma \to 0$ at $y\to y_n$. For a given $y$, the
photon emission angle increases with the increase of $n$. The
electron scattering angle is small,
\begin {equation}
\theta_e \approx {y\theta_\gamma\over 1-y}\leq {2n\, \over
\sqrt{1+\xi^2}}\, {\omega\over m}\, .
 \label{65}
\end {equation}

 \begin{figure}[!ht]
 \begin{center}
\includegraphics[width=0.47\textwidth]{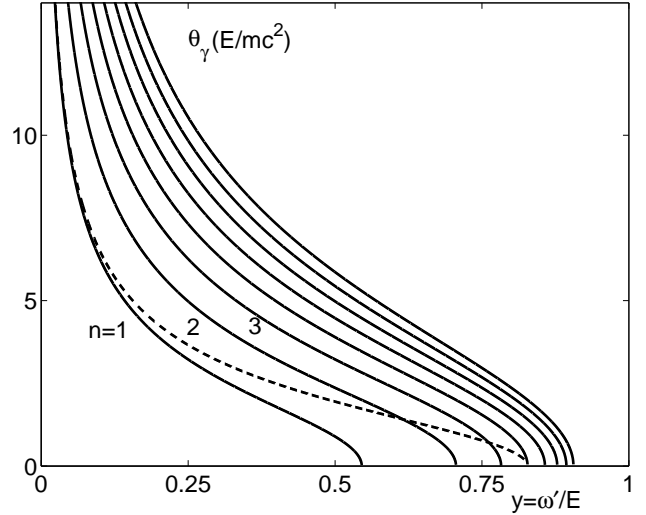}
 \end{center}
\caption{Photon scattering angles $\theta_\gamma$ for different
harmonics $n$ versus the final photon energy $\omega'$ at $x=4.8$
and $\xi^2=3$. The dashed curve corresponds to $\xi^2=0$.}
 \label{f1}
 \end{figure}

Let us consider the azimuthal dependence of the detected Stokes
parameters $\xi'_j$ for the final photon. These parameters are
defined with respect to the $x^{\prime}y^{\prime}z^{\prime}$-axes
which are fixed to the scattering plane. The $x'$-axis is the same
as for the initial photon and perpendicular to the scattering
plane:
\begin{equation}
x' \;\parallel\; {\bf k} \times {\bf k}';
 \label{2.6a}
\end{equation}
the $y'$-axis are in that plane:
\begin{equation}
y' \;\parallel\; {\bf k}' \times ({\bf k}\times {\bf k}')\,.
  \label{2.8a}
\end{equation}
At small emission angles of the final photon $\theta_\gamma \ll
1$, the final photon moves almost along the direction of the
$z$-axis and ${\bf k} \times {\bf k}'$ azimuth is approximately
equal to $\varphi + (3\pi /2)$. Let $\check{\xi}'_j$ be the
detected Stokes parameters for the final photons, fixed to the
$xyz$-axes of the collider system. These Stokes parameters are
connected with $\xi'_j$ by the following relations:
 \bea
\xi'_1 &\approx& -\check{\xi}'_1 \cos{2 \varphi}+ \check{\xi}'_3
\sin{2 \varphi}\,, \; \; \xi'_2 = \check{\xi}' _2 \,,
 \nn
 \\
\xi'_3 &\approx& -\check{\xi}'_3 \cos{2 \varphi}- \check{\xi}'_1
\sin{2 \varphi}
 \label{2.10}\,.
 \eea

Inserting these relations into the effective cross section, we
find the Stokes parameters resulting from the scattering process
itself and fixed to the $xyz$-axes of the collider system:
 \bea
\check{\xi}^{\,(f)}_1 &\approx& -{F_1\over F_0} \cos{2 \varphi}-
{F_3\over F_0} \sin{2 \varphi}\,, \; \; \check{\xi}^{\,(f)}_2 =
{F_2\over F_0} \,,
 \nn
 \\
\check{\xi}^{\,(f)}_3 &\approx& -{F_3\over F_0} \cos{2 \varphi}+
{F_1\over F_0} \sin{2 \varphi}\,.
 \label{2.10f}
 \eea
Note that (\ref{2.10}) and (\ref{2.10f}) become exact in the limit
$y \to y_n$.

It is not difficult to check that, for the considered case, the
angle $\Delta \theta$ between vector $\mbox{\boldmath$\nu$}_3$ and
${\bf n}'_3$ is very small:
 \be
\Delta \theta \approx |\mbox{\boldmath$\nu$}_{3\perp}-{\bf
n}'_{3\perp}| \approx {m \theta_e \over 2 E'} \leq {n\over
\sqrt{1+\xi^2}}\,{\omega\over E'}\, .
 \ee
This means that the invariants $\zeta_j'$ in (\ref{31}) almost
coincide with the projections
$\mbox{\boldmath$\zeta$}'\,\mbox{\boldmath$\nu$}_j$, defined in
(\ref{55}),
 \bea
\zeta_1'&\approx& \zeta'_{\perp}\sin{(\varphi-\beta')}\,,\;\;
\zeta'_2\approx \zeta'_{\perp}\cos{(\varphi-\beta')}\,,
 \nn
 \\
\zeta'_3&\approx& 2\lambda'_e\,,
 \label{56a}
 \eea
and that the exact equation (\ref{61}) for the polarization of the
final electron can be replaced  with a high accuracy by the
approximate equation
 \be
\mbox{\boldmath$\zeta$}^{(f)} \approx \sum_{j=1}^3\,{G_j\over
F_0}\, \mbox{\boldmath$\nu$}_j \,.
 \label{67}
 \ee

Up to now we deal with the head-on collisions of the laser photon
and the initial electron beams, when the collision angle
$\alpha_0$ between the vectors ${\bf p}$ and $(-{\bf k})$ was
equal zero. The detailed consideration of the case $\alpha_0\neq
0$ is given in Appendix B; see also \cite{Grin}. We present here
the summary of such a consideration. If $\alpha_0\neq 0$, the
longitudinal component of the vector ${\bf k}$ becomes
$k_z=-\omega\cos{\alpha_0}$ and it is appeared to be the
transverse (to the momentum ${\bf p}$)  component ${\bf k}_\perp$.
However this transverse component is small, $|{\bf k}_\perp|
\lesssim\omega\ll m$, therefore the transverse momenta of the
final particles almost compensate each other, ${\bf p}_\perp'={\bf
k}_\perp-{\bf k}_\perp'\approx -{\bf k}_\perp'$. The invariants
$x$ and $y$ become (compare with (\ref{63}))
\begin{equation}
x\approx\frac{4E\omega}{m^2}\cos^2{\frac{\alpha_0}{2}},\;\;
y\approx \frac{\omega'}{E}\approx 1-\frac{E'}{E}\,.
 \label{xy0}
\end{equation}
The polarization parameters of the initial and the final electrons
and photons conserve their forms (\ref{51}),
(\ref{56a}),(\ref{20}), (\ref{2.9}), (\ref{2.10}) and
(\ref{2.10f}). Therefore, the whole dependence on $\alpha_0$
enters the effective cross section and the polarizations only via
the quantity $x$ (\ref{xy0}).

\subsection{Averaged polarization of the final particles}

The  polarization of the final photon and electron, averaged over
the azimuthal angle $\varphi$ in the collider system, is important
in many application. To find it, one can use the same method as
for the linear Compton scattering (see~\cite{Khoze},
\cite{GKPS83}, \cite{KPS98}). Let us consider, for definiteness,
the case of the circularly polarized laser photons. Substituting
the exact equations (\ref{51}) for $\zeta_j$ and the approximate
equations (\ref{2.10}), (\ref{56a}) for $\xi'_j$, $\zeta'_j$ into
(\ref{34}), we obtain the explicit dependence of the effective
cross section (\ref{33}) on $\varphi$. Then we take into account
that the terms
 \bea
\sum ^3_{j=1}\, F_j^{(n)}\xi'_j&\approx&
\left(-F_{1}^{(n)}\,\cos{2\varphi}
-F_{3}^{(n)}\,\sin{2\varphi}\right)\,\check{\xi}_1'
 \label{68d}
 \\
&+& F_2^{(n)}\check{\xi}'_2+ \left(F_{1}^{(n)}\,\sin{2\varphi}
-F_{3}^{(n)}\,\cos{2\varphi}\right)\,\check{\xi}_3'
 \nn
 \eea
after the averaging over $\varphi$ become equal to
 \be
\sum ^3_{j=1}\,\langle F_j^{(n)}\xi'_j \rangle \approx \langle
F_2^{(n)} \rangle\, \check{\xi}'_2
 \ee
where
 \bea
\langle F_{2}^{(n)} \rangle&=& \left({1\over 1-y} +1-y\right)\,
c_n h_n P_c
 \nn
 \\
&+& y\left( {2-y\over 1-y}\,f_n- {s^2_n\over 1+\xi^2}\,g_n\right)
\zeta_3\,.
  \eea
After a similar averaging of the terms $\sum_j G_j^{(n)}
\zeta_j'$, we obtain
  \bea
&&{ d\sigma^{(n)}
(\mbox{\boldmath$\zeta$},\,\mbox{\boldmath$\xi$},\,
\mbox{\boldmath$\zeta$}^{\prime},\,\mbox{\boldmath$\xi$}^{\prime})\over
dy} \approx {\pi r_e^2\over 2x}\;\left( \langle F_0^{(n)} \rangle
+  \langle F_2^{(n)} \rangle\, \check{\xi}'_2  \right.
 \nn
 \\
&&\left.+ G_\perp^{(n)} \mbox{\boldmath$\zeta$}_\perp
\mbox{\boldmath$\zeta$}'_\perp + G_\parallel^{(n)} \zeta_3' +
\sum^3_{i,j=1} \langle
H_{ij}^{(n)}\,\zeta^{\prime}_i\,\xi^{\prime}_j \rangle \right)\,,
  \label{26a}
 \eea
where $ \langle F_0^{(n)} \rangle$, $G_\perp^{(n)}$ and
$G_\parallel^{(n)}$ are given in (\ref{54c}), (\ref{48}) and
(\ref{48a}), respectively. According to the usual rules, we obtain
from this equation the averaged polarization of the final
particles for the case of the circularly polarized laser photons:
 \bea
\langle \check{\xi}_{1}^{(n)(f)} \rangle &\approx& 0\,,\; \langle
\check{\xi}_{2}^{(n)(f)} \rangle \approx {\langle F_{2}^{(n)}
\rangle \over \langle F_{0}^{(n)} \rangle}\,,\;\; \langle
\check{\xi}_{3}^{(n)(f)} \rangle \approx 0\,,
 \nn
 \\
\langle \mbox{\boldmath$\zeta$}^{(n)(f)}_{\perp} \rangle & \approx
& {G_{\perp}^{(n)} \over \langle F_{0}^{(n)} \rangle}\,
\mbox{\boldmath$\zeta$}_\perp\,,\;\; \langle \zeta_{3}^{(n)(f)}
\rangle \approx  { G_{\parallel}^{(n)}\over \langle
F_{0}^{(n)}\rangle }\,.
 \label{67b}
 \eea

In a similar way we obtain the averaged polarization of the final
particles for the case of the linearly polarized laser
photons\footnote{To perform such an averaging, we take into
account that the functions $\tilde{f}_n$, $\tilde{g}_n$ and
$\tilde{h}_n$ are even functions of the variable $\cos{\varphi}$.
Therefore, $\langle \tilde{g}_n \sin{\varphi} \rangle =0$, since
the averaged function is odd under the replacement $\varphi \to -
\varphi$, and  $\langle \tilde{g}_n \cos{\varphi} \rangle =\langle
\tilde{h}_n \cos{\varphi} \rangle =0$, since the averaged function
is odd under the replacement $\varphi \to \pi - \varphi$.}:
  \bea
\langle \check{\xi}_{1}^{(n)(f)} \rangle &\approx& 0\,,\;\;
\langle \check{\xi}_{2}^{(n)(f)}\rangle \approx {\langle
F_{2}^{(n)} \rangle \over \langle F_{0}^{(n)} \rangle}\,,
 \nn
 \\
\langle \check{\xi}_{3}^{(n)(f)} \rangle &\approx& {1\over\langle
F_{0}^{(n)} \rangle}\,\left[ 2\langle\tilde{f}_n\rangle-
(2+2c_n-s_n^2\Delta)\,\langle \tilde{g}_n\rangle \right.
 \nn
 \\
&+&\left. (1+c_n)^2\, \langle\tilde{g}_n\cos{2\varphi}\rangle
\right],
 \label{67c}
 \\
\langle \mbox{\boldmath$\zeta$}^{(n)(f)}_{\perp} \rangle &\approx&
 {\langle \tilde{G}_{\perp}^{(n)}\rangle \over \langle F_{0}^{(n)}
\rangle}\, \mbox{\boldmath$\zeta$}_\perp\,,\;\; \langle
\zeta_{3}^{(n)(f)} \rangle \approx  {\langle
\tilde{G}_{\parallel}^{(n)} \rangle \over \langle
F_{0}^{(n)}\rangle } \,,
 \nn
 \eea
where $\langle F_{0}^{(n)}\rangle$, $\tilde{G}_{\perp}^{(n)}$,
$\tilde{G}_{\parallel}^{(n)}$ are given in (\ref{54cl}),
(\ref{2.25a}), (\ref{2.25ab}) and
 \be
 \langle F_{2}^{(n)}\rangle=y\left( {2-y\over 1-y}\,\langle
 \tilde{f}_n \rangle- {s^2_n\over
1+\xi^2}\, \langle\tilde{g}_n \rangle \right)\, \zeta_3\,.
 \label{67f}
 \ee

Note that the averaged Stokes parameters of the final photon do
not depend on $\mbox{\boldmath$\zeta$}_\perp$ and that the
averaged polarization vector of the final electron is not equal
zero only if $P_c \neq 0$ or if $\mbox{\boldmath$\zeta$} \neq 0$.
These properties are similar to those in linear Compton
scattering.

\section{Limiting cases}

In this section we consider several limiting cases in which
description of the non-linear Compton scattering is essentially
simplified.

{\it (i). The case of $\;\xi^2 \to 0$.} At small $\xi^2$ all
harmonics with $n > 1$ disappear due to the properties (\ref{42}),
(\ref{44}), (\ref{2.20}), (\ref{2.21}) and we have
 \be
d\sigma^{(n)}(\mbox{\boldmath$\zeta$},\,\mbox{\boldmath$\xi$},\,
\mbox{\boldmath$\zeta$}^{\prime},\,\mbox{\boldmath$\xi$}^{\prime})
 \propto \xi^{2(n-1)}
\;\; \mbox{ at } \;\; \xi^2\to 0\,.
 \label{c1}
 \ee
We have checked that in this limit our expression for the
effective cross section of the first harmonic, $d\sigma^{(1)}$,
coincides with the result known for the linear Compton effect; see
Appendix A.

{\it (ii). The case of $\;y \to 0$ for arbitrary $x$ and $\xi^2$.}
This limit corresponds to forward scattering ($\theta_e=0\,,\;
\theta_\gamma =\pi$). We expect that in this limit an electron as
well as a photon does not change its polarization. Indeed, at
small $y$ all harmonics with $n
> 1$ disappear due to the properties (\ref{42}), (\ref{44}) and
(\ref{2.20}), (\ref{2.21}), i.e.
 \be
d\sigma^{(n)}(\mbox{\boldmath$\zeta$},\,\mbox{\boldmath$\xi$},\,
\mbox{\boldmath$\zeta$}^{\prime},\,\mbox{\boldmath$\xi$}^{\prime})
 \propto y^{n-1}
\;\; \mbox{ at } \;\; y\to 0\,,
 \label{c1a}
 \ee
and we found that
 \be
F_0=2\,,\;\; F_j= 2 \xi_j\,,\;\; G_j = 2 \zeta_j\,,\;\; H_{ij} = 2
\zeta_i\,\xi_j\,,
 \ee
therefore,
\be
{d\sigma (\mbox{\boldmath$\zeta$},\,\mbox{\boldmath$\xi$},\,
\mbox{\boldmath$\zeta$}^{\prime},\,\mbox{\boldmath$\xi$}^{\prime})
\over dy\, d\varphi} = {r_e^2\over 2x}\, (1+
\mbox{\boldmath$\zeta$}
\mbox{\boldmath$\zeta$}^{\prime})(1+\mbox{\boldmath$\xi$}
\mbox{\boldmath$\xi$}^{\prime})
 \label{c1b}
 \ee
It follows from this equation that
 \be
\mbox{\boldmath$\zeta$}^{(f)}=\mbox{\boldmath$\zeta$}\,,\;\;
\mbox{\boldmath$\xi$}^{(f)}=\mbox{\boldmath$\xi$}\,,
 \label{c1be}
 \ee
in accordance with our expectation.

Note that for the case of circularly polarized laser photons the
equations, similar to (\ref{c1be}), hold also for each harmonic
separately:
$\mbox{\boldmath$\zeta$}^{(n)(f)}=\mbox{\boldmath$\zeta$}\,,\;\;
\mbox{\boldmath$\xi$}^{(n)(f)} =\mbox{\boldmath$\xi$}$.

{\it (iii). The case of $\;x \to 0$ for arbitrary $\xi^2$}, which
corresponds to the classical limit. In this limit an electron does
not change its polarization. For a certain harmonic with $n\, x
\ll 1$ one has $y\leq y_n < nx \ll 1$ and
 \be
G_j^{(n)}= F_0^{(n)}\,\zeta_j\,,\;\; H_{ij}^{(n)} = \zeta_i\,
F_j^{(n)}\,,
 \ee
therefore, in this limit the effective cross section both for the
circularly and linearly polarized laser photons has the form
 \bea
&&{d\sigma^{(n)}(\mbox{\boldmath$\zeta$},\,\mbox{\boldmath$\xi$},\,
\mbox{\boldmath$\zeta$}^{\prime},\,\mbox{\boldmath$\xi$}^{\prime})
\over dy\, d\varphi}
 \label{c2a}
 \\
&&  = {r_e^2\over 4x}\, \left( F_0^{(n)} + \sum_{j=1}^3 \,
F_j^{(n)}\, \xi_j' \right)\, \left(1+ \sum_{i=1}^3 \, \zeta_i
\zeta'_i \right)\,.
 \nn
 \eea
As a consequence, the polarization of the final electron for a
given harmonic $n$ is
 \be
\mbox{\boldmath$\zeta$}^{(n)(f)}= \sum_{i=1}^3\, \zeta_i\, {\bf
n}'_i\,,
 \label{polx}
 \ee
where the unit vectors ${\bf n}'_i$ are defined in (\ref{31}). To
show that $\mbox{\boldmath$\zeta$}^{(n)(f)}$ coincides with
$\mbox{\boldmath$\zeta$}$, one has to demonstrate that the unit
vectors ${\bf n}'_i$ coincide with the unit vectors ${\bf n}_i$
from (\ref{28}). This can be easily seen, for example, in the
frame of reference in which ${\bf p} =0$ and in which the photon
energies are small in the considered limit, $\omega' < n \omega
\ll m$, as well as the momentum of the final electron, $|{\bf p}'|
\ll m$. As a result,
 \be
\mbox{\boldmath$\zeta$}^{(n)(f)}=\mbox{\boldmath$\zeta$}\,,
 \ee
in accordance with our expectation.

{\it (iv). The case of $\;\theta_\gamma \to 0$ for the circularly
polarized laser photons, $P_c=\pm 1$.} At small $\theta_\gamma$
the energy of the final photon is close to its maximum, $y\to
y_n$, and the functions (\ref{40}) become equal to
 \bea
f_n&=&g_n=h_n= \gamma_n\,\left(y_n-y\right)^{n-1}\,,
 \\
\gamma_n&=&(n^n/n!)^2\, \left[ {\xi^2\over  y_n (1-y_n)(1+\xi^2)}
\right]^{n-1}\,.
 \nn
 \eea
Therefore, all harmonics with $n>1$ disappear\footnote{The
vanishing for the strict backward scattering of all harmonics with
$n \geq 3$ is due to the conservation of $z$ component of the
total angular momentum $J_z$ in this limit: the initial value of
$J_z= \lambda_e-n\,\lambda_L$ can not be equal for $n \geq 3$ to
the final value of $J_z=\lambda_e' +\lambda'_\gamma$ (here
$\lambda_e$ $(\lambda'_e)$ is the helicity of the initial (final)
electron and $\lambda_L$ $(\lambda'_\gamma)$ is the helicity of
the laser (final) photon). For $n=2$ this argument does not help,
since $J_z=\pm 3/2$ can be realized for the initial and final
states. The vanishing of the second harmonics for the strict
backward scattering is a specific feature of the process, related
to the facts that $J_z=\pm 3/2$ can be realized only if
$\lambda'_\gamma =- \lambda_L$ and that the polarization vector
$e^{(\lambda'_\gamma)*} =- e^{(\lambda_L)}$ is orthogonal not only
to the 4-vectors $k,\, k'$ but to the 4-vectors $p,\,p'$ and
$e^{(\lambda_L)}$ as well.} and only the photons of the first
harmonic can be emitted along the direction of the initial
electron beam:
 \bea
{d\sigma^{(n)}(\mbox{\boldmath$\zeta$},\,\mbox{\boldmath$\xi$},\,
\mbox{\boldmath$\zeta$}^{\prime},\,\mbox{\boldmath$\xi$}^{\prime})
\over dy d\varphi} &=&{r^2_e\over 4x}\, F^{(n)}\propto f_n
 \propto (y_n -y)^{n-1}
 \nn
  \\
&&\mbox{ at } \;\; y\to y_n\,,
 \label{c1d}
 \eea
with
 \bea
{F^{(n)}\over f_n}&=& {2-2y_n+y_n^2\over 1-y_n}\, (1+\zeta_3
\zeta_3')(1-P_c\xi_2')
 \label{Fn}
 \\
& -&{y_n(2-y_n)\over 1-y_n}\, (\zeta_3+ \zeta_3')(P_c-\xi_2')
 \nn
 \\
&+& 2 (\mbox{\boldmath$\zeta$}_\perp
\mbox{\boldmath$\zeta$}'_\perp)(1-P_c\xi_2')\,.
 \nn
 \eea
In this limit, the final photons are circularly polarized,
 \be
\xi_1^{(n)(f)}=0\,,\;\; \xi_2^{(n)(f)}=-P_c\,,\;\;
\xi_3^{(n)(f)}=0\,,
 \label{ppol}
 \ee
and the polarization of the final electrons is
 \bea
\mbox{\boldmath$\zeta$}_\perp^{(n)(f)}&=& {2(1-y_n)\over 2
-(2y_n-y_n^2)(1+\zeta_3P_c)}\,\mbox{\boldmath$\zeta$}_\perp\,,
 \label{elpol}
 \\
\zeta_3^{(n)(f)}&=& {2\zeta_3 -(2y_n-y_n^2)(\zeta_3+P_c)\over 2
-(2y_n-y_n^2)(1+\zeta_3P_c)}\,,
 \nn
 \eea
in particular, at $\zeta_3=\pm 1$ the helicity of the electron is
conserved in the non-linear Compton scattering,
$\zeta_3^{(n)(f)}(y\to y_n) \to \zeta_3$. Note that the function
$F^{(n)}$ does not depend on the azimuthal angle $\varphi$ and
(\ref{ppol}) and (\ref{elpol}) are in accordance with the results
(\ref{67b}) for the polarizations averaged over $\varphi$.

Let us consider in more detail this limit for the first harmonic.
The spectrum of the fist harmonic,
 \bea
{d\sigma^{(1)}(\mbox{\boldmath$\zeta$},\,\mbox{\boldmath$\xi$})
\over dy} &=&{2\pi r^2_e\over x(1-y_1)}
[2-(2y_1-y_1^2)(1+\zeta_3P_c)]
 \nn
 \\
&&\mbox{ at } \; y\to y_1\,,
 \label{specir}
 \eea
depends on the non-linearity parameter $ \xi^2$ only via
 $$
y_1= {x\over x+1+\xi^2}\,.
 $$
For the case of ``good polarization'', $\zeta_3P_c=-1$, the first
harmonic has a peak at $y\to y_1$ of the value
 \bea
{d\sigma^{(1)}(\mbox{\boldmath$\zeta$},\,\mbox{\boldmath$\xi$})
\over dy} &=&{4\pi r^2_e\over x}\,\left(1+{x\over 1+\xi^2} \right)
 \nn
 \\
&&\mbox{ at } \;\; y\to y_1\;\; \mbox{ and }\;\; \zeta_3P_c=-1\,.
 \label{maxspecir}
 \eea

At large $x\gg 1+\xi^2$ the quantity $y_1$ tends to $1$,
 \be
 1-y_1 ={1+\xi^2 \over x+1+\xi^2} \ll 1\,,
 \ee
and the function $F^{(1)}$ becomes large
 \bea
F^{(1)}&=& {1\over 1-y_1}\,\left[ (1+\zeta_3
\zeta_3')(1-P_c\xi_2') \right.
 \nn
 \\
&-& \left. (\zeta_3+ \zeta_3')(P_c-\xi_2')\right]\,.
 \eea
Therefore, integrating the effective cross section for the first
harmonic, $d\sigma^{(1)} \propto 1/(1-y)$, near its maximum, i.e.
in the region
 \be
 {1+\xi^2 \over x} \ll 1-y \ll 1\,,
 \ee
we find (with logarithmic accuracy) the total cross section
 \bea
\sigma(\mbox{\boldmath$\zeta$},\,\mbox{\boldmath$\xi$},\,
\mbox{\boldmath$\zeta$}^{\prime},\,\mbox{\boldmath$\xi$}^{\prime})
&=& {\pi r_e^2\over 2x} \,\left[ (1+\zeta_3 \zeta_3')(1-P_c\xi_2')
\right.
 \nn
 \\
&-& \left. (\zeta_3+ \zeta_3')(P_c-\xi_2')\right]\, \ln{x\over
1+\xi^2}\,.
 \label{total}
 \eea
If one sums this expression over spin states of the final
particles, one arrives at the result for the cross section
 \be
\sigma(\mbox{\boldmath$\zeta$},\,\mbox{\boldmath$\xi$}) = {2\pi
r_e^2\over x} \,\left(1-\zeta_3 P_c \right)\, \ln{x\over
1+\xi^2}\,,
 \ee
which is in accordance with that obtained by a different method
in~\cite{BKMS}.

{\it (v). The case of $\;\theta_\gamma \to 0$ for the linearly
polarized laser photons.} At $y\to y_n$, the argument $a$ tends to
zero, while the argument $b$ tends to the nonzero constant, $b\to
n \xi^2/[2(1+\xi^2)]$; the function $A_0(n,\,a,\,b)\propto
\sqrt{y_n-y}$ vanishes for odd $n$ and $A_0(n,\,a,\,b)\to
(-1)^{n/2} J_{n/2}(b)$ for even $n$. As a result, the spectra and
the polarizations are different in this limit for odd and even
harmonics.

{\it Odd harmonics}. The spectrum in this case,
 \bea
{d\sigma^{(n)}(\mbox{\boldmath$\zeta$},\,\mbox{\boldmath$\xi$})
\over dy} &=&{2\pi r^2_e\over
x(1-y_n)}\,\left(2-2y_n+y_n^2\right)
 \label{speclin}
 \\
&\times&\left[J_{(n-1)/2}(b)-J_{(n+1)/2}(b) \right]^2 \; \mbox{ at
} \;\; y\to y_n\,,
 \nn
 \eea
depends on the non-linearity parameter $\xi^2$ via $y_n$ and via
the argument $b$, but the spectrum does not depend on the initial
electron polarization. In this limit, the final photons have
circular polarization, proportional to $\zeta_3$, and linear
polarization along the direction of the linear polarization of the
laser photons,
 \bea
\check{\xi}_1^{(n)(f)}&=&0\,,\;\;\check{\xi}_2^{(n)(f)}={2y_n-y^2_n\over
2-2y_n+y_n^2}\,\zeta_3\,,
 \nn
 \\
\check{\xi}_3^{(n)(f)}&=&{2(1-y_n)\over 2-2y_n+y_n^2}\,.
 \label{ppollin}
 \eea
The polarization of the final electrons in this case is
 \bea
\mbox{\boldmath$\zeta$}_\perp^{(n)(f)}&=& {2(1-y_n)\over 2
-2y_n+y_n^2}\,\mbox{\boldmath$\zeta$}_\perp\,,
 \label{elpollin}
 \\
\zeta_3^{(n)(f)}&=& \zeta_3\,,
 \nn
 \eea
in particular, the final electrons have the same mean helicity as
that for the initial one.

{\it Even harmonics}. The spectrum reads
 \bea
{d\sigma^{(n)}(\mbox{\boldmath$\zeta$},\,\mbox{\boldmath$\xi$})
\over dy} &=&{2\pi r^2_e\over x(1-y_n)}\,{2y_n^2\over \xi^2}\,
\left[J_{n/2}(b)\right]^2
 \nn
 \\
 && \mbox{ at } \;\; y\to y_n\,.
 \label{specline}
 \eea
In this limit the final photons have only circular polarization
equals to $\zeta_3$, and the final electrons have mean helicity
opposite to that for the initial one:
 \bea
&&\check{\xi}_1^{(n)(f)}=0\,,\;\;
\check{\xi}_2^{(n)(f)}=\zeta_3\,, \;\; \check{\xi}_3^{(n)(f)}=0\,,
 \label{ppolline}
 \\
&&\mbox{\boldmath$\zeta$}_\perp^{(n)(f)}=0\,,\;\;
\zeta_3^{(n)(f)}= -\zeta_3\,.
 \nn
 \eea
Note that (\ref{ppollin}), (\ref{elpollin}) and (\ref{ppolline})
are in accordance with the results (\ref{67c}) for the
polarizations averaged over $\varphi$.

\section{Numerical results related to $\gamma \gamma$ and
$\gamma e$ colliders}

In this section we present some examples which illustrate the
dependence of the differential cross sections and the
polarizations, obtained in the previous sections, on the final
photon energy. We restrict ourselves to properties of the
high-energy photon beam which are of most importance for the
future $\gamma \gamma$ and $\gamma e$ colliders. The used
parameters are close to those in the TESLA project\cite{TESLA}. In
particular,
 $$
 x\approx 4E\omega/m^2=4.8\,,
 $$
and the non-linearity parameter $\xi^2$ (\ref{5}) is chosen either
the same as in the TESLA project, $\xi^2=0.3$, or larger by one
order of magnitude, $\xi^2=3$, just to illustrate the tendencies.
In figures below we use notation
 $$
\sigma_0=\pi r_e^2 \approx 2.5\cdot 10^{-25} \;\;\mbox{cm}^2\,.
 $$

\begin{figure}[b]
\includegraphics[width=0.47\textwidth]{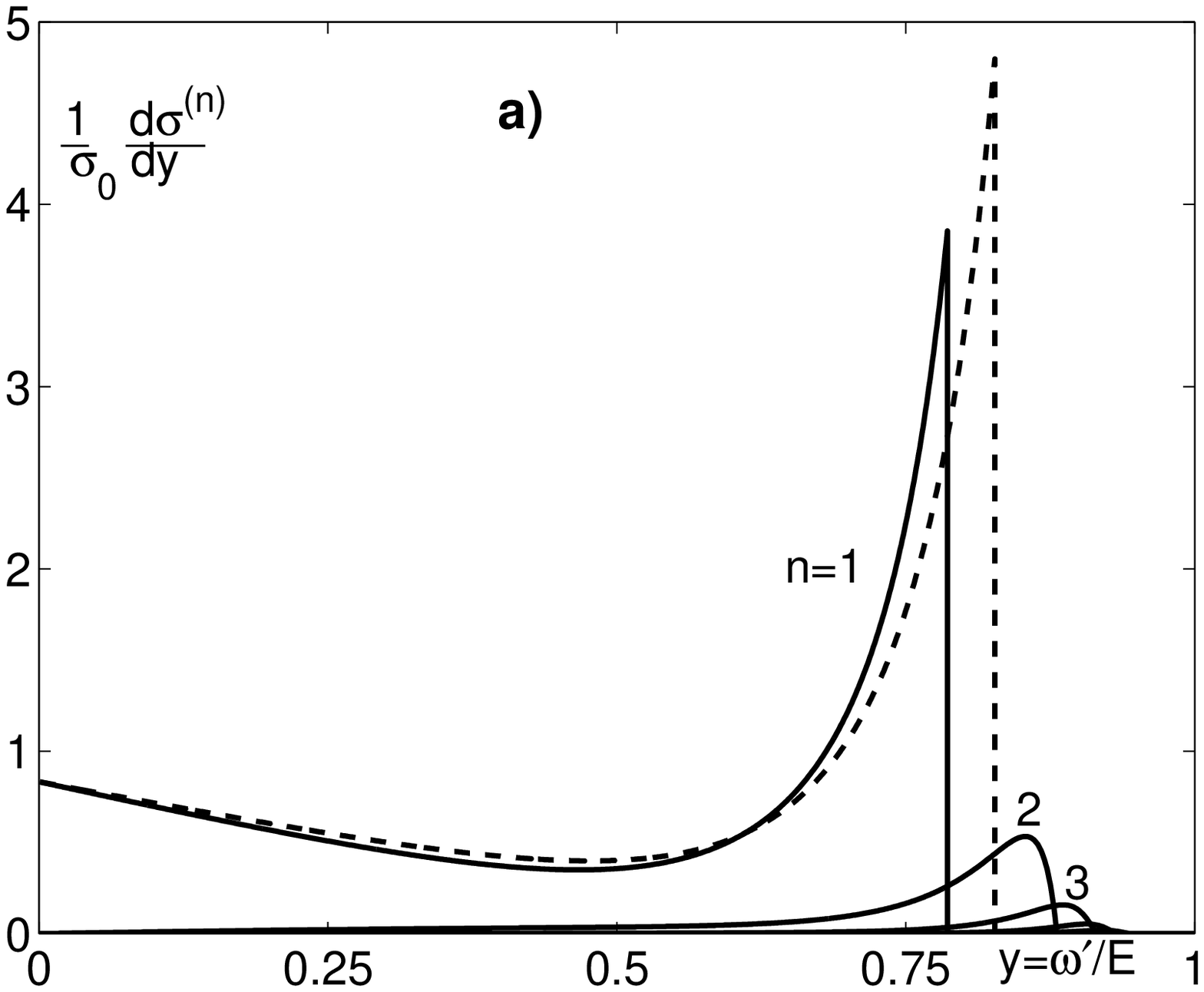}
\includegraphics[width=0.47\textwidth]{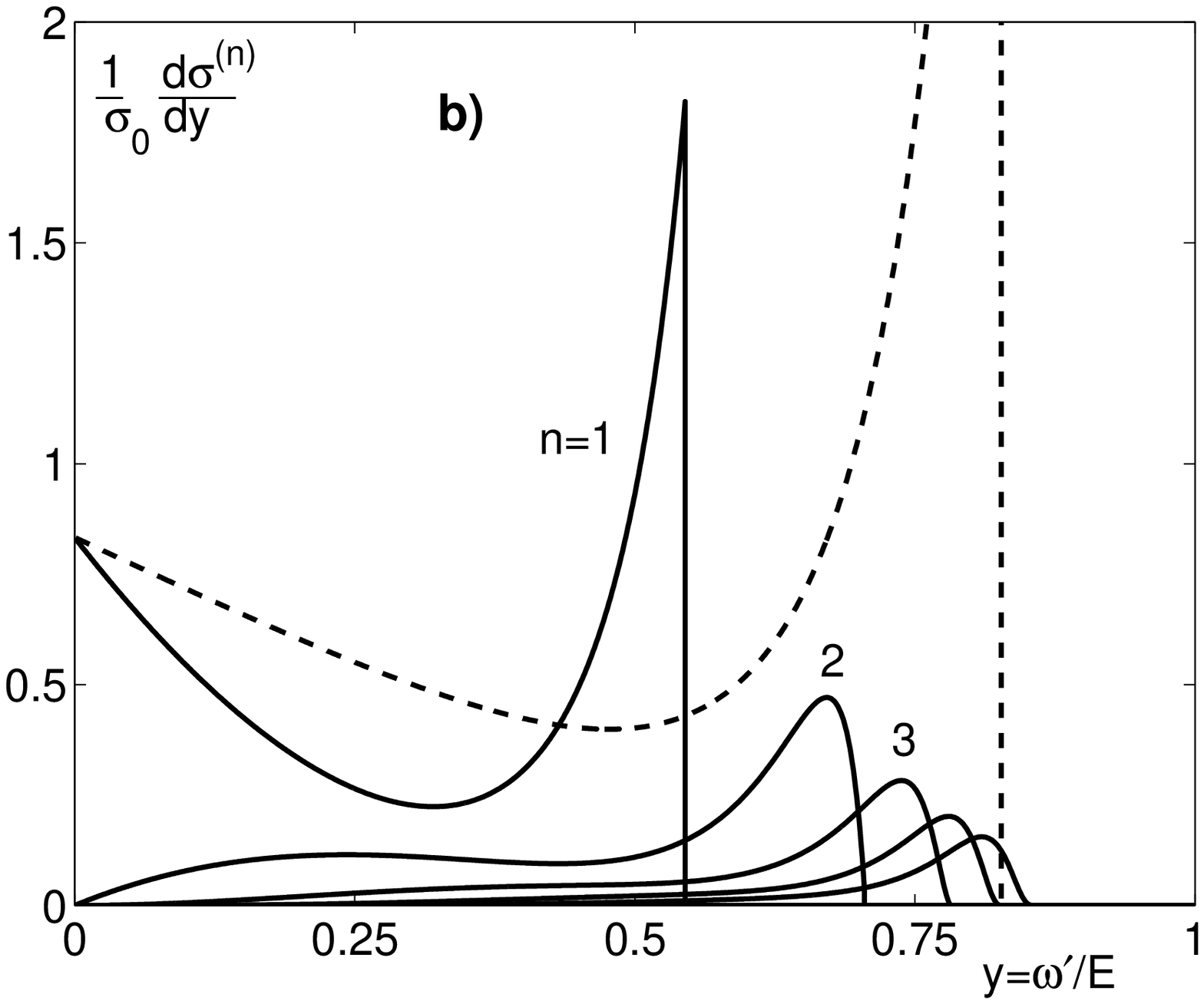}
 \caption{Energy spectra of final photons for different
harmonics $n$ at $\zeta_3\,P_c =-1$ and: {\bf (a)} $\xi^2=0.3$;
$\;$ {\bf (b)} $\xi^2=3$. The dashed curves correspond to
$\xi^2=0$. }
 \label{f2}
\end{figure}
\vspace{7mm}
\begin{figure}[b]
 \begin{center}
\includegraphics[width=8cm]{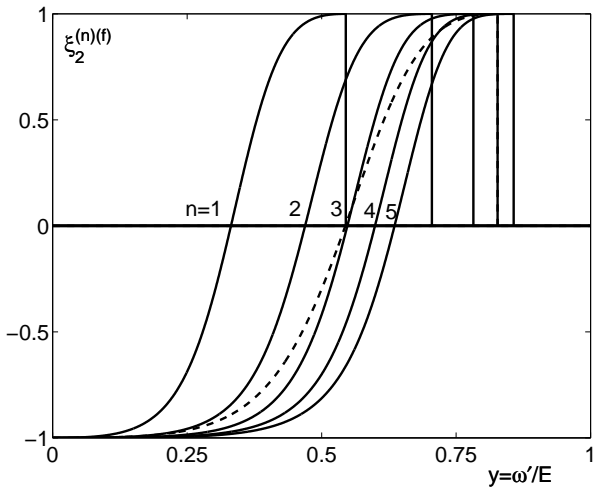}
 \end{center}
\caption{Mean helicity of the final photons for different
harmonics $n$ versus the final photon energy $\omega'$ at $P_c
=-1,\;\zeta_3=1$ and $\xi^2=3$. The dashed curve corresponds to
$\xi^2=0$.}
 \label{f3}
\end{figure}

{\it The case of the circularly polarized laser photons (Figs.
\ref{f2}, \ref{f3}, \ref{f4}, \ref{f5}).}

The spectra of the few first harmonics are shown in Fig. 2 for the
case of ``a good polarization'', when helicities of the laser
photon and the initial electron are opposite, $\zeta_3P_c=-1$. At
a small intensity of the laser wave ($\xi^2 = 0.3$, Fig. 2a) the
main contribution is given by photons of the first harmonic and
the probability for generation of the higher harmonics is small.
However, with the growth of the non-linearity parameter ($\xi^2 =
3$, Fig. 2b), the maximum energy for the first harmonic decreases
and the peak of this harmonic at $y=y_1$ (see Eq.
(\ref{maxspecir})) decreases as well. As for the higher harmonics,
with the rise of $\xi^2$ (Fig. 2b) we see an increase of the yield
of photons with energies higher than the maximum energy of the
first harmonic. As a result, the total spectrum becomes
considerable wider.

\begin{figure}[]
\includegraphics[width=0.51\textwidth]{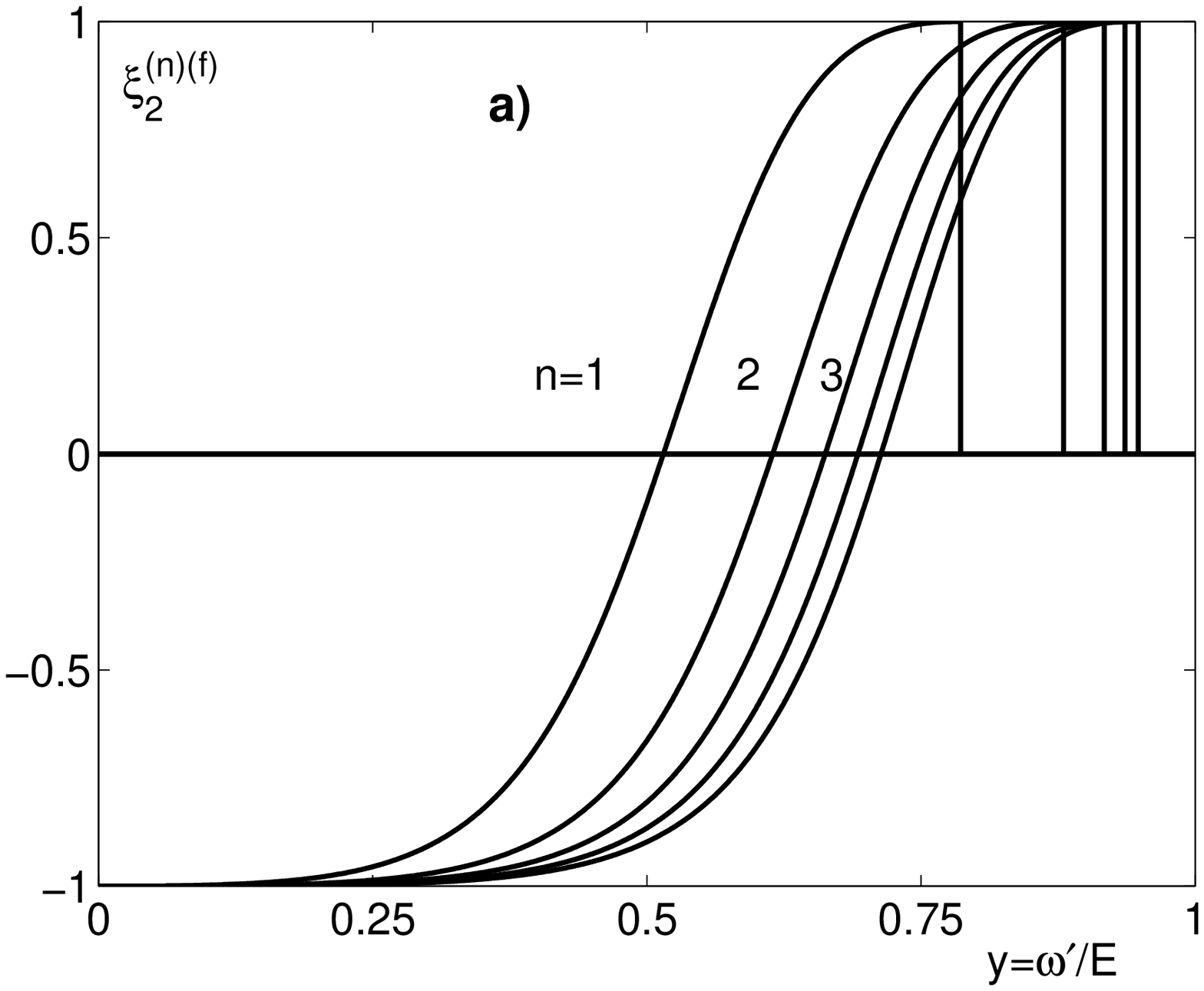}
\includegraphics[width=0.47\textwidth]{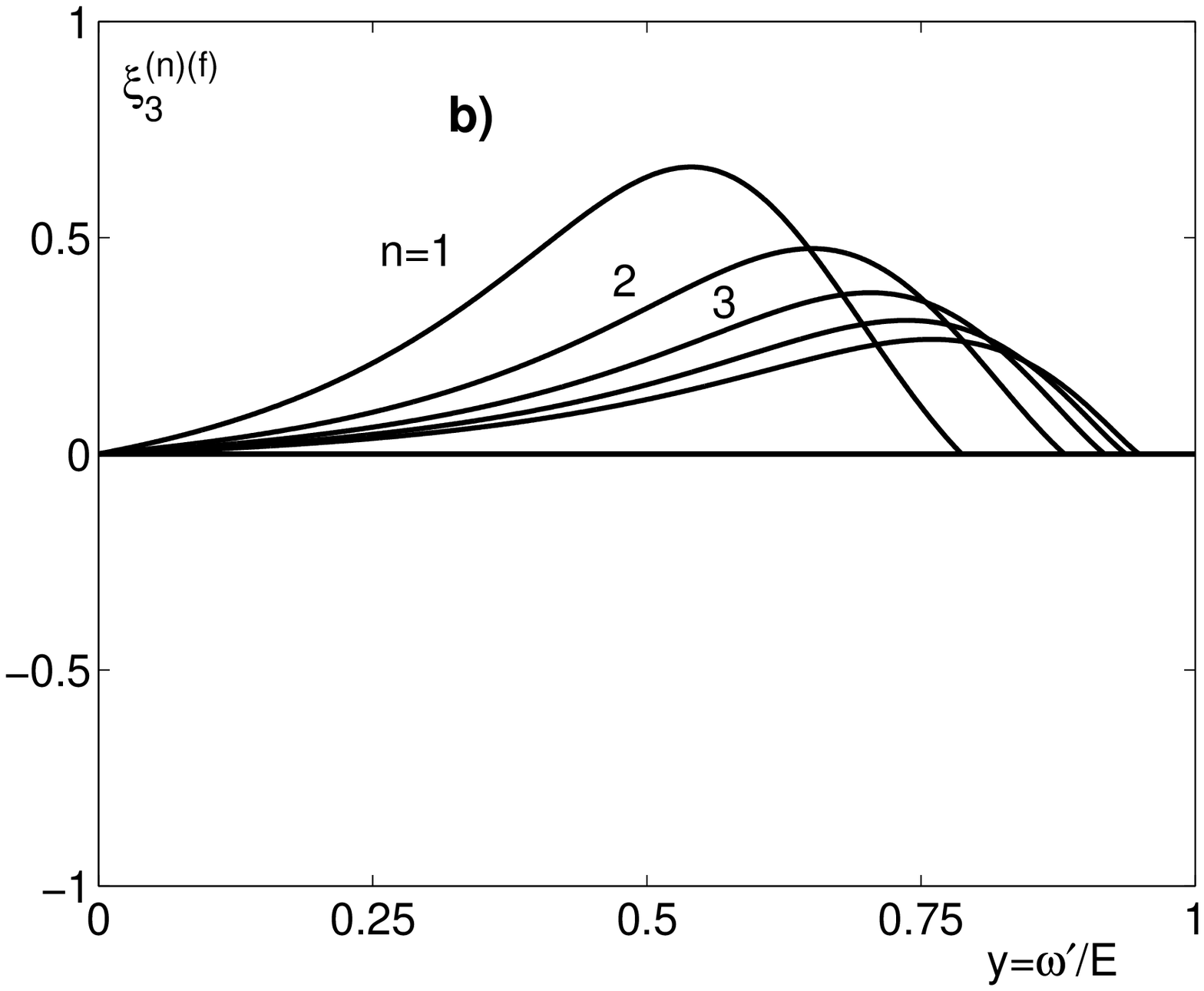}
 \caption{The Stokes parameters of final photons for different
harmonics $n$ versus the final photon energy $\omega'$ at
$\zeta_3=1,\;P_c =-1$, $\xi^2=0.3$: {\bf (a)} the mean helicity
$\xi_2^{(n)(f)}$; $\;$ {\bf (b)} the linear polarization
$\xi_3^{(n)(f)}$ transverse to the scattering plane. The Stokes
 parameter obeys $\xi_1^{(n)(f)}=0$ in this case.}
 \label{f4}
\end{figure}
\begin{figure}[]
\includegraphics[width=0.47\textwidth]{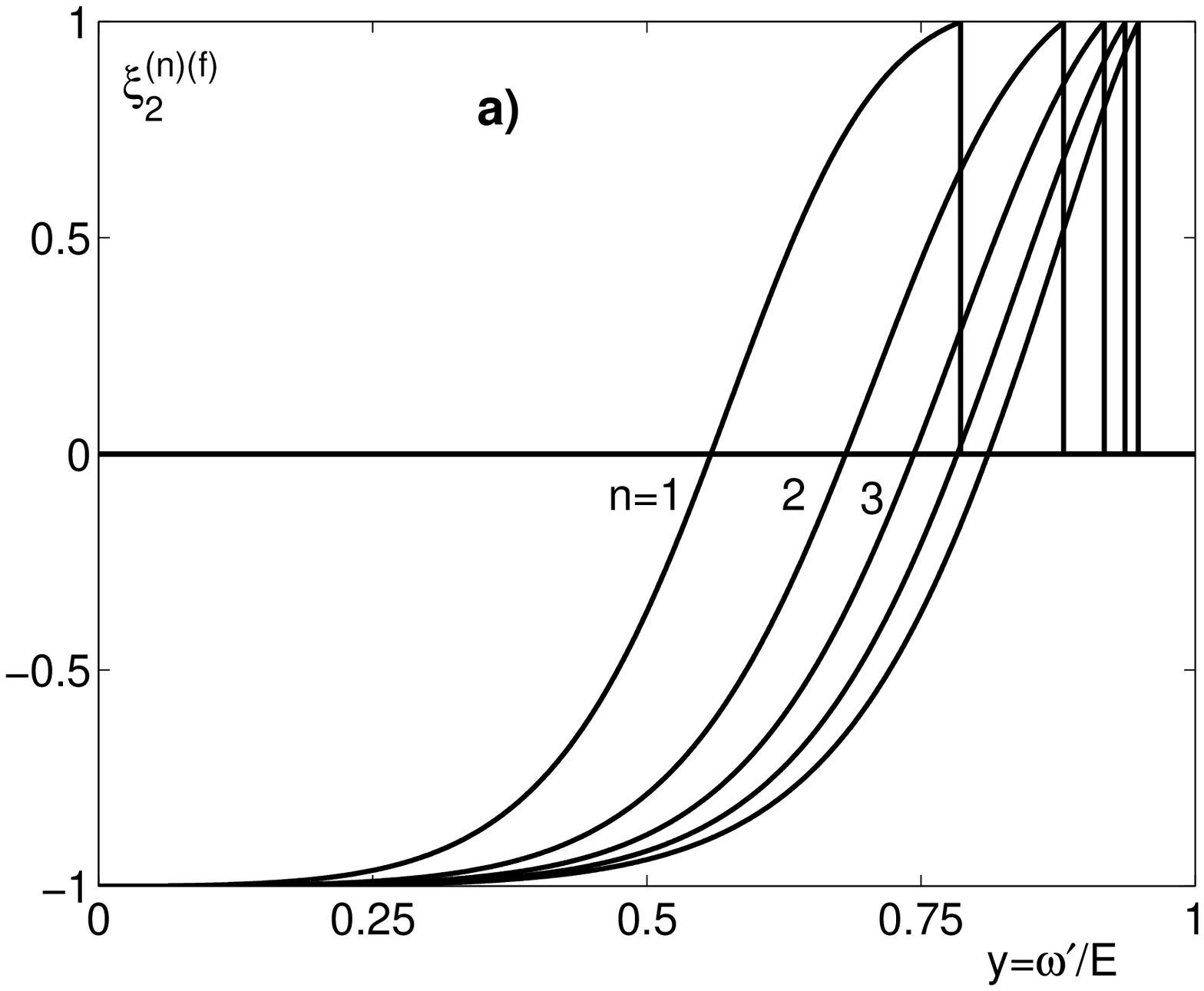}
\includegraphics[width=0.47\textwidth]{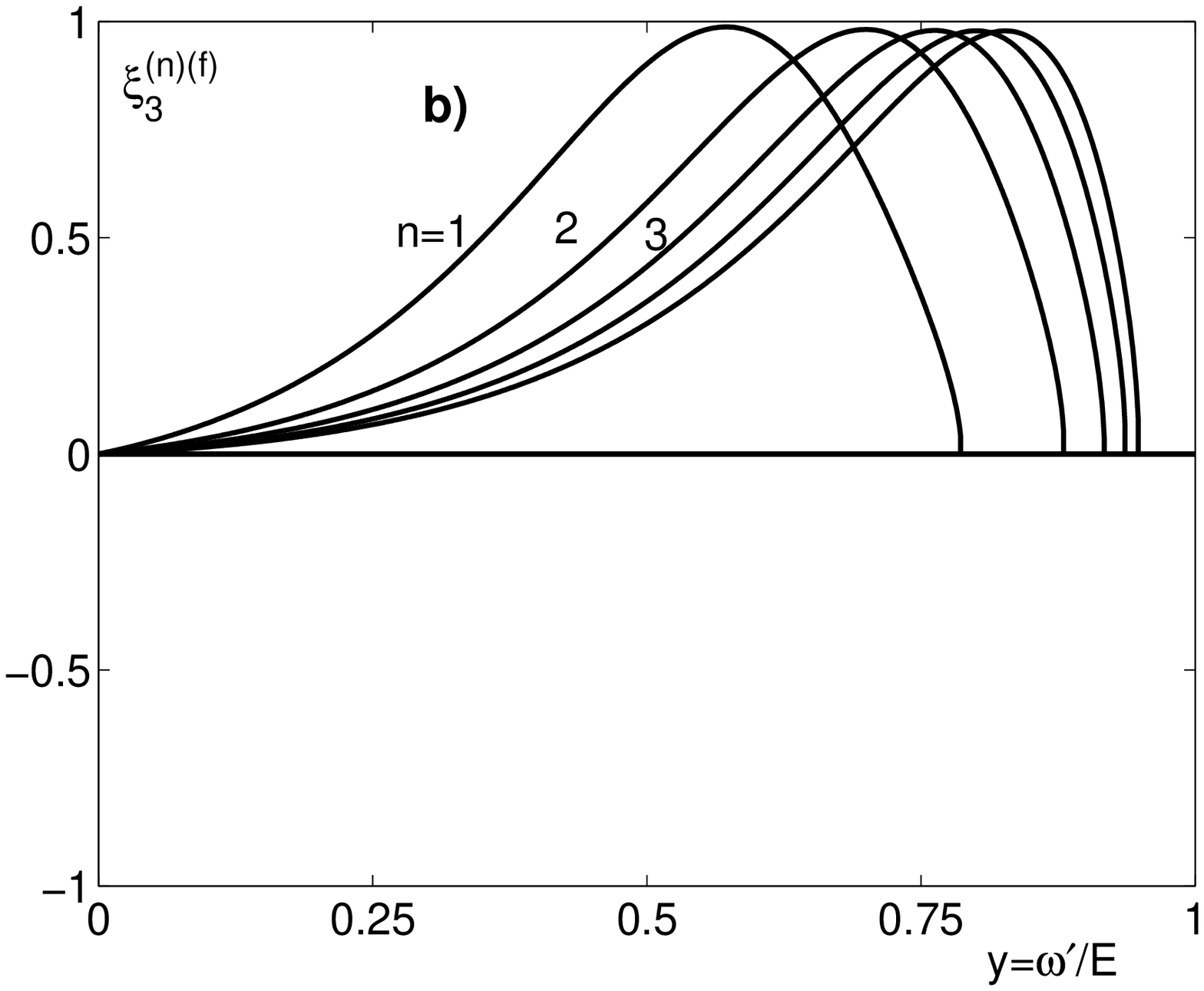}
\caption{The same as in Fig. 4, but for $\zeta_1=0,\; \zeta_2=0.7$
and $\zeta_3=0.7$.}
 \label{f5}
\end{figure}

The mean helicities of the final photons for the few first
harmonics, $\xi_2^{(n)(f)}$, are shown in Fig. 3 and Figs. 4a, 5a.
In Fig. 3 we compare the two cases $\xi^2=3$ and $\xi^2=0$. Each
harmonic is 100\% polarized near its maximum $y=y_n$, therefore,
the whole $\gamma$ beam at $x>0.5$ has a high degree of circular
polarization. However, for linear Compton scattering, the final
photon in the region of high energies,  $y\geq 0.8$, is almost 100
\% circularly polarized, while for the non-linear case $\xi^2=3$,
only harmonics with $n\geq 4$ contribute in this energy range. But
the cross sections corresponding to these harmonics are small (see
Fig. 2b).

There is an interesting feature related to the linear polarization
of the final photons; see Figs 4b and 5b.  Such a polarization is
described by the Stokes parameters $\xi_1^{(n)(f)}$ and
$\xi_3^{(n)(f)}$. These parameters averaged over  the azimuthal
angle $\varphi$ (see (\ref{67b})) vanish. On the other hand, the
degree of the linear polarization for a given angle $\varphi$ may
be close to 100\%, especially in the case when the initial
electron has a transverse polarization as well (see Fig. 5b). It
is not excluded that this property of our process may result in a
sizable dependence of the  luminosities of $\gamma \gamma$
collisions at a photon collider on the mutual orientation
(parallel or perpendicular) of the  linear polarizations of the
colliding photons. Such a dependence is of importance, for
example, for the study of a Higgs boson production at a $\gamma
\gamma$ collider. As far as we know, this possibility has not been
discussed so far even for linear Compton scattering.

{\it The case of the linearly polarized laser photons (Figs.
\ref{f6}, \ref{f7}, \ref{f8}).}

The spectra of the first few harmonics for this case are shown in
Fig. 6. They differ considerably from those for the case of the
circularly polarized laser photons shown in Fig. 2. First of all,
in the considered case the spectra do not depend on the
polarization of the initial electrons. The maximum of the first
harmonic at $y=y_1$ now is about two times smaller than that on
Fig. 2. Besides, the harmonics with $n>1$ do not vanish at $y=y_n$
contrary to such harmonics on Fig. 2.

For the first few harmonics the mean helicities of the final
photons, $\xi_2^{(n)(f)}$, are shown at $\zeta_3=1$  in Fig. 7a
for $\xi^2=0.3$ and in Fig. 7b for $\xi^2=3$. We would like to
direct attention to the surprising fact that each harmonic is
almost 100\% circularly polarized near the high-energy part of the
spectrum. The curves on Fig. 7 are given for the azimuthal angle
$\varphi=0$, for other values of $\varphi$ these curves would look
a bit different, but not too much.

The degree of linear polarization of the final photons is not
large in the high-energy part of the spectrum, but it becomes
rather high in the middle and the low part of the spectrum; see
Fig. 8. Certainly, the direction of this polarization depends on
the azimuthal angle $\varphi$, as the result the linear
polarization averaged over $\varphi$ is substantially smaller, see
Fig. 9, than one on Fig. 8. Nevertheless, the non-trivial effects,
related to the high degree of linear polarization present at a
certain $\varphi$, do exist. For example, it was shown
\cite{PST2003} that for linear Compton scattering this feature
leads to an important effect for the luminosity of $\gamma\gamma$
collisions.

\begin{figure}[!htb]
\includegraphics[width=8cm]{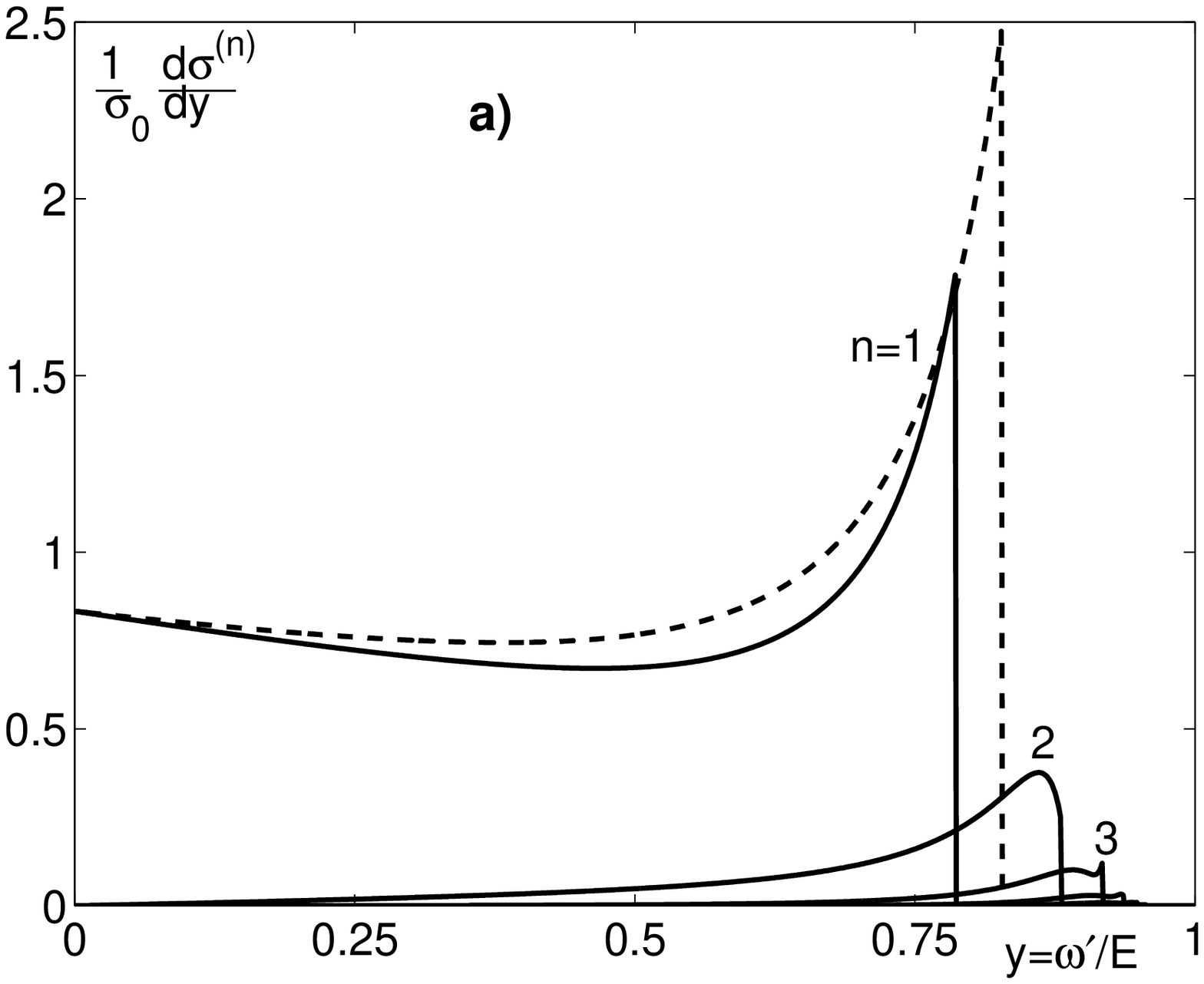}
\includegraphics[width=8cm]{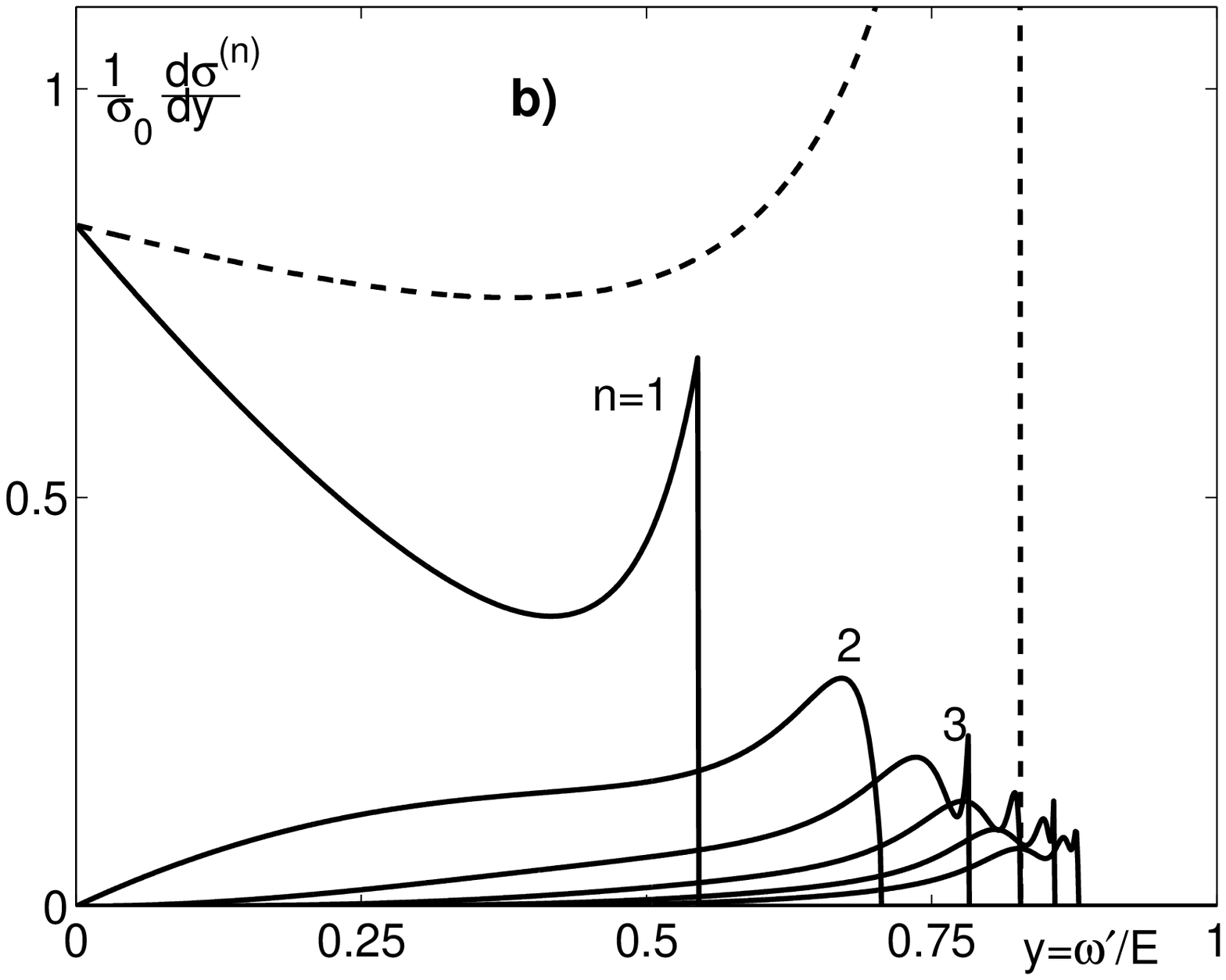}
\caption{Energy spectra of final photons for different harmonics
$n$ in the case of the linearly polarized laser photons and: {\bf
(a)} $\xi^2=0.3$; $\;$ {\bf (b)} $\xi^2=3$. The dashed curves
correspond to $\xi^2=0$.}
 \label{f6}
\end{figure}

\begin{figure}[b]
\includegraphics[width=0.47\textwidth]{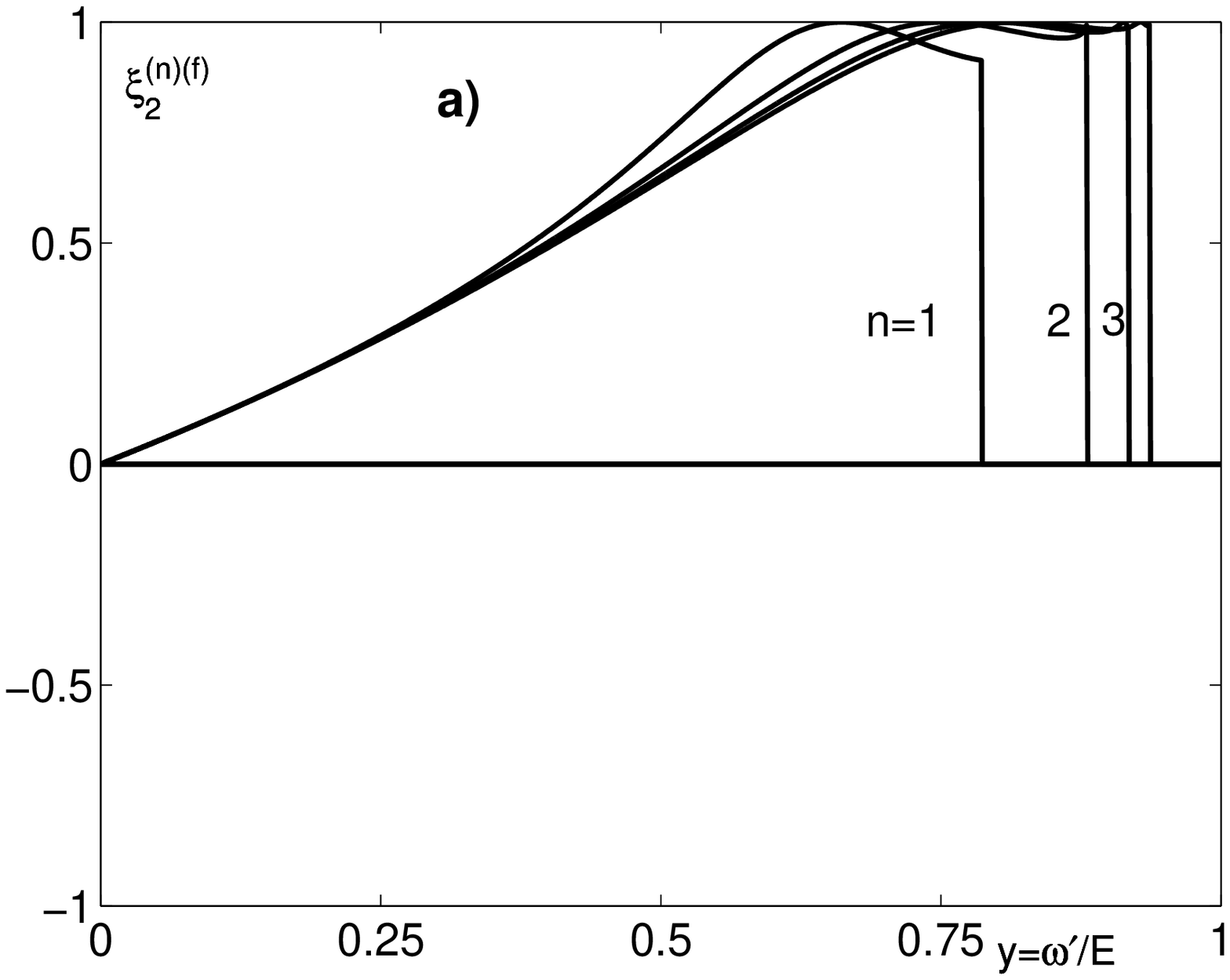}
\includegraphics[width=0.47\textwidth]{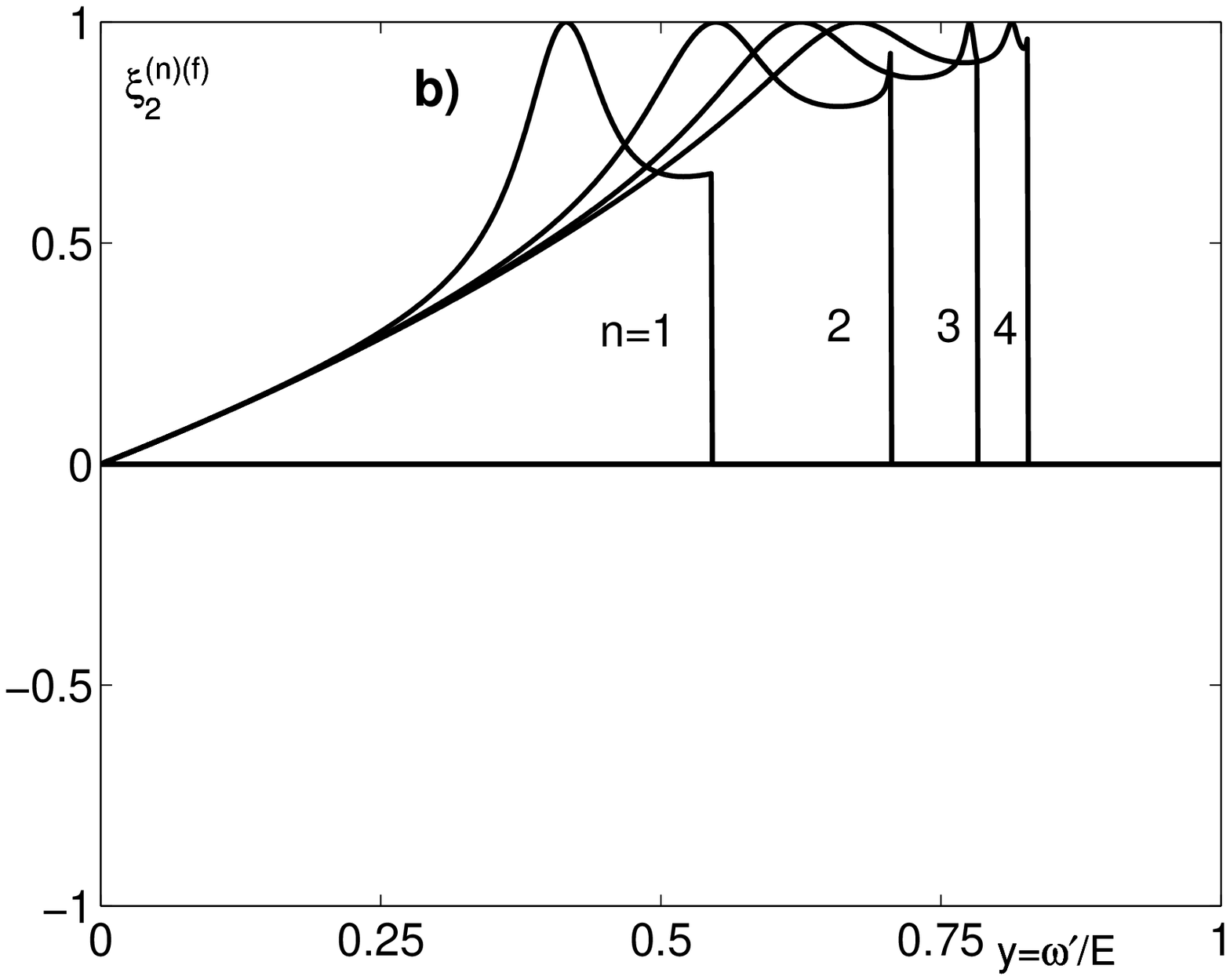}
\caption{The mean helicity $\xi_2^{(n)(f)}$ of the final photons
for different harmonics $n$ versus the final photon energy
$\omega'$ at $\zeta_3=1$ and: {\bf (a)} $\xi^2=0.3$; $\;$ {\bf
(b)} $\xi^2=3$. The laser photons are linearly polarized. The
scattering plane is parallel to the direction of the laser photon
polarization.}
 \label{f7}
\end{figure}

\begin{figure}[b]
\includegraphics[width=0.47\textwidth]{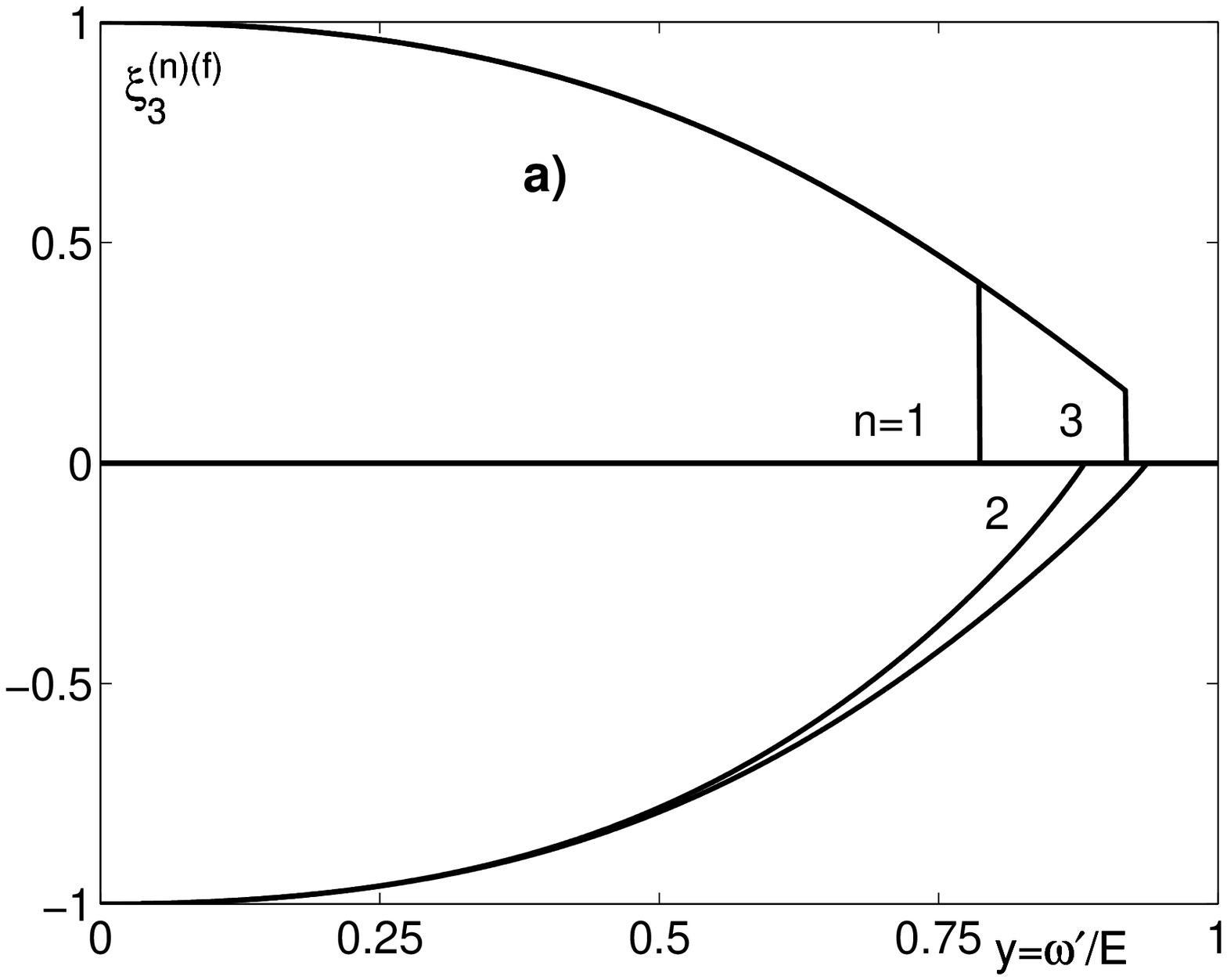}
\includegraphics[width=0.47\textwidth]{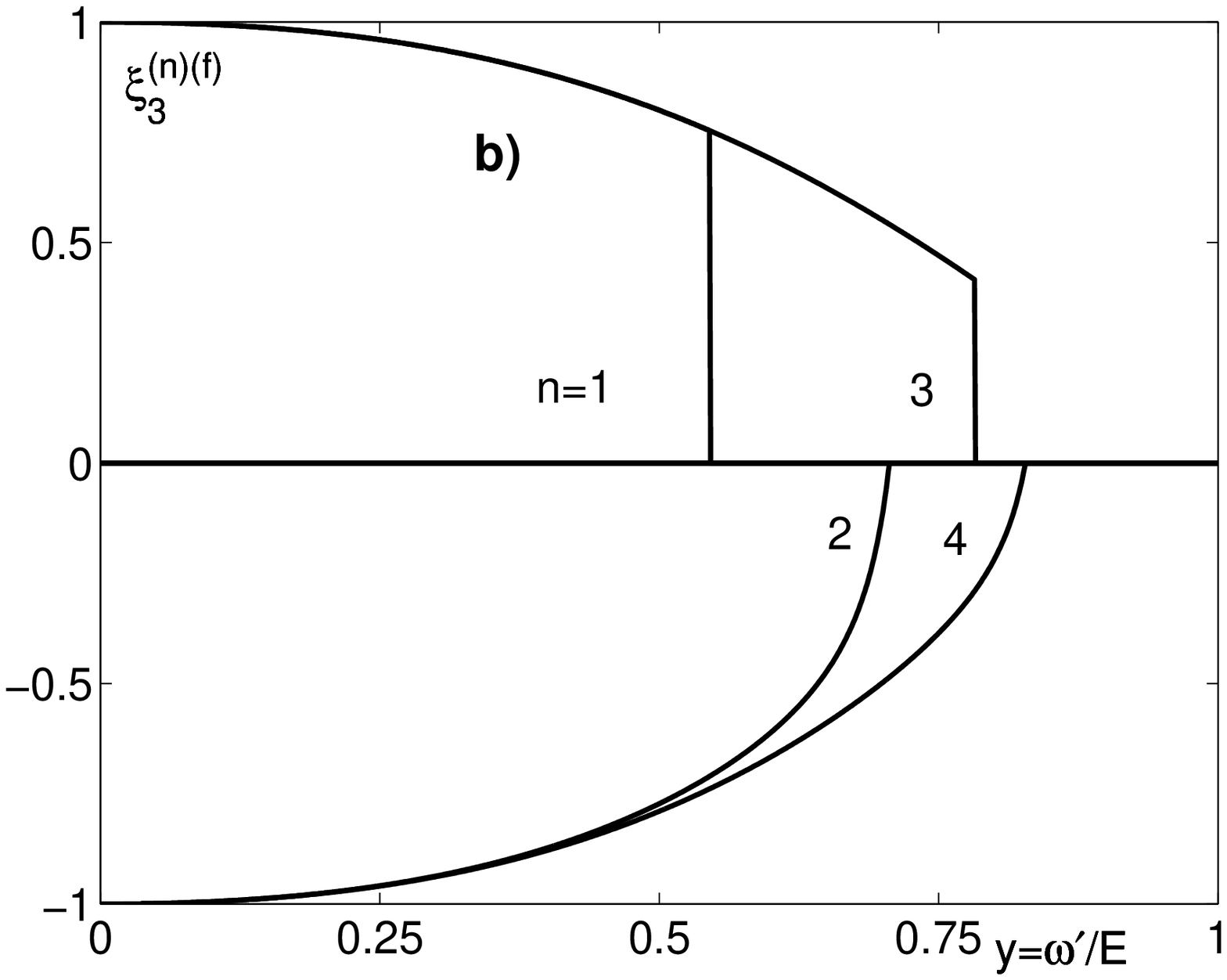}
\caption{The Stokes parameters $\xi_3^{(n)(f)}$ of the final
photons for different harmonics $n$ versus the final photon energy
$\omega'$ at: {\bf (a)} $\xi^2=0.3$; $\;$ {\bf (b)} $\xi^2=3$. The
laser photons are linearly polarized. The scattering plane is
perpendicular to the direction of the laser photon polarization.
The Stokes parameter obeys $\xi_1^{(n)(f)}=0$ in this case. }
 \label{f8}
\end{figure}

\begin{figure}[b]
\includegraphics[width=0.47\textwidth]{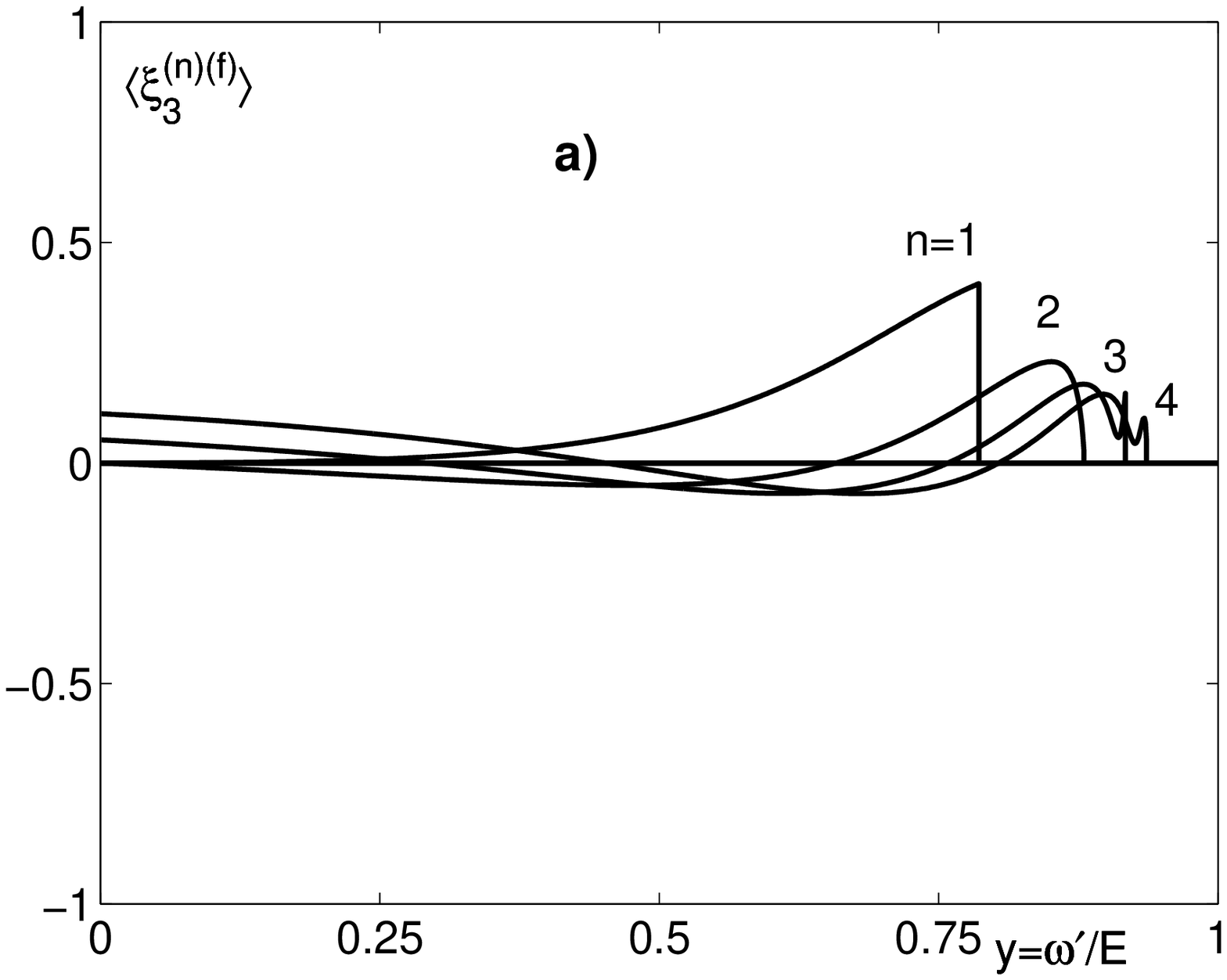}
\includegraphics[width=0.47\textwidth]{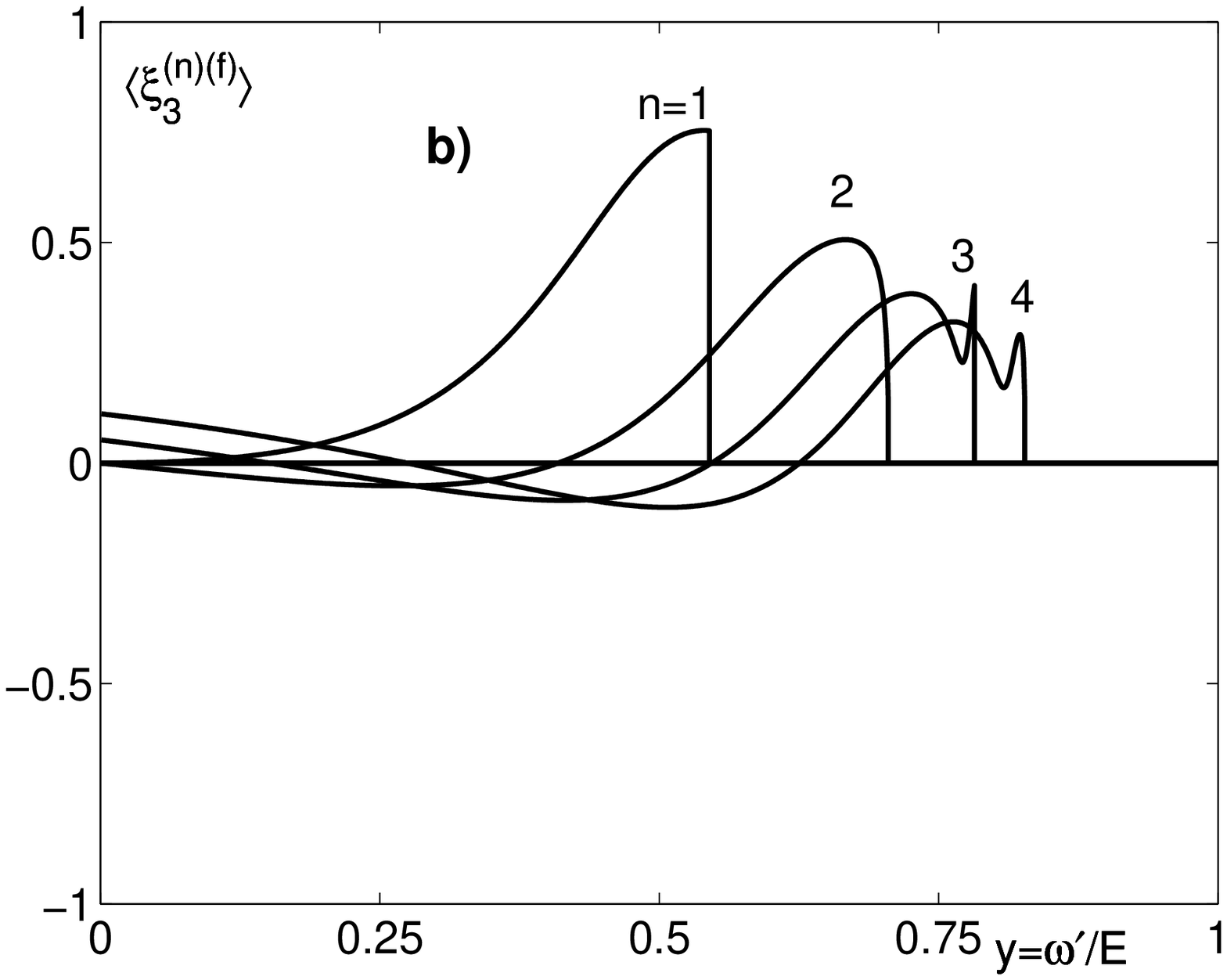}
\caption{The Stokes parameters $\langle\xi_3^{(n)(f)}\rangle$ of
the final photons for different harmonics $n$, averaged over the
azimuthal angle $\varphi$, versus the final photon energy
$\omega'$ at: {\bf (a)} $\xi^2=0.3$; $\;$ {\bf (b)} $\xi^2=3$. The
laser photons are linearly polarized.  The averaged Stokes
parameter $\langle \xi_1^{(n)(f)}\rangle=0$ in this case. }
 \label{f9}
\end{figure}

\section{Summary and comparison with other papers}

Our main results are given by (\ref{45})--(\ref{49}) for
circularly and by (\ref{2.23})--(\ref{2.26}) for linearly
polarized laser beam. They are expressed in terms of the 16
functions $F_0,\; F_j,\; G_j$ and $H_{ij}$ with $i,\,j = 1\div 3$,
which describe completely the polarization properties of the
non-linear Compton scattering in a compact invariant form. The
functions $F_0$ and $F_j$ enter the total cross section
(\ref{35}), (\ref{54}), the differential cross sections
(\ref{53}), (\ref{54}) and the Stokes parameters of the final
photons (\ref{36}). The polarization of the final electrons is
described by functions $G_j$, which enter the polarization vector
$\mbox{\boldmath$\zeta$}^{(f)}$ given by the exact equations
(\ref{38}), (\ref{61}) and the approximate equations (\ref{67}).

Besides, we considered the kinematics and the approximate formulae
relevant for the problem of $e\to \gamma$ conversion at the
$\gamma \gamma$ and $\gamma e$ colliders. In particular, we
discuss the polarization of the final photons and electrons
averaged over the azimuthal angle $\varphi$ (see  (\ref{67b}),
(\ref{67c})) and the spectra and the polarization of the final
particles in the limit of small and large energies of the final
photons (Sect. 5). For circularly polarized laser photons and for
large values of the parameter $x \gg 1+\xi^2$, we obtain (with
logarithmic accuracy) the simple analytical expression for the
total cross section (\ref{total}). In Sect. 6 we present the
numerical results and discuss several interesting features in the
final photon polarization, which may be useful for the analysis of
the luminosity of $\gamma \gamma$ collisions.

Let us compare our results with those obtained earlier.

{\it Circular polarization of laser photons}. In the literature we
found  the results which can be compared with our functions $F_0,
F_2$ and $G_j$, namely, in~\cite{NR-review} (the function $F_0$
only at $\zeta_j=0$), in~\cite{GS,Tsai,Yokoya} (the functions
$F_0$ and $F_2$ only at $\zeta_1=\zeta_2=0$) and in~\cite{BPM}
(the function $F_0$ for arbitrary $\zeta_j$). For these cases our
results coincide with the above mentioned ones. For polarization
of the final electron, our results differ slightly from those
in~\cite{Yokoya,BPM}. Namely, the function $G_2$ at
$\zeta_1=\zeta_2=0$ given in~\cite{Yokoya} and the functions $G_j$
for arbitrary $\zeta_j$ obtained in~\cite{BPM} coincide with ours.
However, the polarization vector $\mbox{\boldmath$\zeta$}^{(f)}$
in the collider system is obtained in papers~\cite{Yokoya,BPM}
only in the approximate form equivalent to our approximate
equation (\ref{67}).

{\it Linear polarization of laser photons}. For this case we found
in the literature the expressions which can be compared with ours
functions $F_0, F_1,\, F_2,\, F_3$. The results in
~\cite{NR-review} are in agreement with our function $F_0$
(\ref{2.23}), and the results in~\cite{GR} are in agreement with
our functions $F_j$  (\ref{2.24}).

The correlation of the final particles' polarizations are
described by the functions $H_{ij}$ given in (\ref{49}),
(\ref{2.26}). We did not find in the literature any results  for
these functions both for the case of the circular and the linear
polarization.

\section*{Acknowledgments}

We are grateful to I.~Ginzburg, M.~Galynskii, A.~Milshtein,
S.~Polityko and V.~Telnov for useful discussions. This work is
partly supported by INTAS (code 00-00679) and RFBR (code
02-02-17884 and 03-02-17734); D.Yu.I. acknowledges the support of
Alexander von Humboldt Foundation.

\section*{Appendix A: Limit of the weak laser field}

At $\xi^2\to 0$, the cross section (\ref{33}) has the form
 \bea
&&d\sigma (\mbox{\boldmath$\zeta$},\,\mbox{\boldmath$\xi$},\,
\mbox{\boldmath$\zeta$}^{\prime},\,\mbox{\boldmath$\xi$}^{\prime})
 = {r_e^2\over 4x}\;F\;d\Gamma\,,
 \label{a1}
 \\
&&d\Gamma = \delta (p+k-p'-k')\;{d^3k'\over \omega'}{d^3p'\over
E'}\,,
 \nn
 \eea
where
\begin{equation}
F=F_0+\sum ^3_{j=1}\left( F_j\xi'_j\; + \;G_j
\zeta^{\prime}_j\right) +
\sum^3_{i,j=1}H_{ij}\,\zeta^{\prime}_i\,\xi^{\prime}_j \,.
 \label{a2}
\end{equation}
To compare it with the cross section for linear Compton
scattering, we should take into account that the Stokes parameters
of the initial photon have the values (\ref{20}) for the circular
polarization and the values (\ref{2.9}) for the linear
polarization. Besides, our invariants $c_1$, $s_1$, $r_1$ and the
auxiliary functions (\ref{2.22}) transform at $\xi^2\to 0$ to
 \bea
c_1&\to& c=1- 2r\,,\;\; s_1\to s =2\sqrt{r(1-r)}\,,
 \nn
 \\ r_1&\to& r={y\over x(1-y)}\,,
  \label{a3}
  \\
X_1 &\to& -c\,,\;\; Y_1 \to c(1-\xi_3)\,,\;\; V_1 \to -\xi_3\,.
 \nn
 \eea
Our functions
 \bea
F_0&=&{1\over 1-y}+1-y-s^2(1-\xi_3)
 \nn
 \\
&-&y\left(s\, \zeta_2 -{2-y\over 1-y}c\, \zeta_3\right) \xi_2\,,
 \nn\\
F_1&=&2c \xi_1+{y\over 1-y} s \, \zeta_1\,\xi_2\,,
 \label{a4}
 \\
F_2&=&\left({1\over 1-y} +1-y\right)\, c\,\xi_2- {ys c}\,\zeta_2
 \nn
 \\
&+& y\left( {2-y\over 1-y}\,- s^2\right) \zeta_3 -ys\,\zeta_1
\xi_1+
 ys\left(c\zeta_2+s\zeta_3 \right)\, \xi_3
\,,
 \nn
 \\
F_3&=& s^2  -{y\over 1-y} s\, \zeta_2\,\xi_2+(1+c^2)\,\xi_3
 \nn
 \eea
coincide with those in~\cite{GKPS83}. Our functions
 \bea
G_1&=&(1+c^2+s^2\xi_3)\zeta_1-{y\over 1-y}s\,\zeta_3\,\xi_1\,,
 \label{a5}
 \\
G_2&=& -{ys\over 1-y} \xi_2 + (1+ c^2+ s^2 \xi_3 )\; \zeta_2 -
{ysc \over 1-y} \, (1- \xi_3 ) \;\zeta_3\,,
 \nn
  \\
G_3 &=& yc\, {2-y \over 1-y} \xi_2+ys\zeta_1\xi_1 + ysc (1
-\xi_3)\; \zeta_2
 \nn
 \\
&+& \left[1+ \left( {1\over 1-y} -y \right)\,(c^2 +s^2 \xi_3 )
\right] \; \zeta_3
 \nn
 \eea
coincide with the functions $\Phi_j$ given by Eqs. (31)
in~\cite{KPS98}. Finally, we check that our functions
 \bea
H_{11}&=&ys\xi_2+{2-2y+y^2\over 1-y}\,c\,\zeta_1\,\xi_1
 \nn
 \\
&+&y\left[ {2-y\over 1-y}\,\xi_3 +s^2(1-\xi_3)\right]\,\zeta_2
-ysc(1-\xi_3)\,\zeta_3\,,
 \nn\\
 H_{21}&=&-\frac{y}{1-y}\left[
s^2+(1-y+c^2)\xi_3\right]\zeta_1
 \nn
 \\
&+& {2-2y+y^2\over 1-y}\,c\xi_1\zeta_2 -ys\xi_1\zeta_3\, ,
 \nonumber \\
H_{31}&=& \frac{y c s}{1-y}\,(1-\xi_3) \zeta_1+\frac{y
s}{1-y}\,\xi_1 \zeta_2 +2c\xi_1\zeta_3 \,,
 \nn
 \\
H_{12}&=&2 c \zeta_1 \xi_2 -{ys\over 1-y}\, \xi_1 \, ,
 \label{2.33} \\
H_{22}&=&-\frac{y c s}{1-y} \,(1-\xi_3)+2 c \,\zeta_2\,\xi_2
-\frac{y s }{1-y} \, \zeta_3\,\xi_2 \,,
 \nn\\
H_{32}&=&\frac{y}{1-y} \left[2-y-s^2(1-\xi_3)\right]+y s\,
 \zeta_2\,\xi_2
  \nn
  \\
&+&\frac{2-2y+y^2}{1-y}c \,  \zeta_3\,\xi_2 \, ,
 \nonumber\\
H_{13}&= &s^2\,\zeta_1+ \left(1+c^2 +\frac{y^2}{1-y}\right)\,
\xi_3 \zeta_1
 \nn
 \\
&-&yc\,{2-y\over 1-y}\,\xi_1\, \zeta_2- ys\,\xi_1\zeta_3\,,
 \nn\\
H_{23}&=&- ys\,\xi_2+y\frac{2-y}{1-y}\,c\,\xi_1\, \zeta_1+
\frac{1-y+y^2}{1-y}\,s^2\,\zeta_2
 \nn\\
 &+&\left(1+c^2
+\frac{y^2c^2}{1-y}\right)\,\xi_3\zeta_2+ysc(1-\xi_3)\,\zeta_3 \,,
 \nonumber
 \\
H_{33}&=&\frac{y s}{1-y}\,\xi_1\, \zeta_1 -
\frac{ysc}{1-y}\,(1-\xi_3)\,\zeta_2
 \nn
 \\
&+& s^2\,\zeta_3+(1+c^2) \,\xi_3\,\zeta_3
 \nn
 \eea
coincide with the corresponding functions in~\cite{Grozin}. Making
such a comparison, one has to take into account that the set of
unit 4-vectors used in our paper (see (\ref{19}), (\ref{23})), and
one used in paper \cite{Grozin} are different. The relations
between our notation and the notation of~\cite{Grozin}, which are
marked below by super-index $G$, are the following:
 \bea
\xi_{1,3}^G &=& -\xi_{1,3}\,,\;\; \xi_{2}^G = \xi_{2}\,,\;\;
\xi_{1,3}^{'G} = -\xi'_{1,3}\,,\;\; \xi_{2}^{'G} = \xi'_{2}\,,
\nn \label{aa}
 \\
\zeta_{1,3}^G& =&\zeta_{3,1}\,,\;\; \zeta_2^G=-\zeta_2\,,
 \\
\zeta_{1}^{'G}& =&c\zeta_{3}'- s \zeta_2'\,,\;\; \zeta_{2}^{'G} =
-c\zeta_{2}'- s \zeta_3'\,,\;\; \zeta_3^{'G} =\zeta_1'\,.
 \nn
  \eea
Besides, in~\cite{Grozin} there is a misprint, namely, the factor
$1/4$ should be inserted in the right-hand-side of the equation,
which is analogous to our (\ref{a1}).

\section*{Appendix B:\\ The case of nonzero beam collision angle}

In this appendix  we consider the case when the beam collision
angle $\alpha_0$ between the vectors ${\bf p}$ and $(-{\bf k})$ is
not equal to zero. We omit systematically terms of the order of
$\omega/m$, $m/E$, $\theta_\gamma$ and $\theta_e$. In this
approximation the vectors (\ref{19}) take the form
 \bea
{\bf e}_\perp^{(1)}&=&{{\bf p}\times {\bf p}'\over |{\bf p}\times
{\bf p}'|}\,,\;\;\; {\bf e}_\perp^{(2)}=-{{\bf k}'_\perp\over
|{\bf k}'_\perp|}\,,
 \nn
 \\
e_0^{(j)}&=&e_z^{(j)}= {{\bf e}_\perp^{(j)}{\bf k}_\perp\over
(1+\cos{\alpha_0})\,\omega}\,,
 \label{B1}
 \\
e^{(i)}\,e^{(j)}&=&- {\bf e}_\perp^{(i)}\,{\bf
e}_\perp^{(j)}=-\delta_{ij}\,,
 \nn
 \eea
i.e. the transverse components ${\bf e}_\perp^{(j)}$ do not differ
from their values at $\alpha_0=0$, and zero and the $z$-components
of $e^{(j)}$ vanish in the limit ${\bf k}_\perp \to 0$. As a
consequence, the vector
 \be
{\bf e}_{{\rm L}\perp}= {\bf e}_\perp^{(1)} \,\sin{\varphi}- {\bf
e}_\perp^{(2)} \,\cos{\varphi} \label{zzz}
 \ee
also does not differ from its value at $\alpha_0=0$. This means
that the parameter $\varphi$ in all the equations of Subsect. 3.3
is the same azimuthal angle of the final photon counted from the
$x$-axis  chosen along the direction of the vector ${\bf e}_{{\rm
L}\perp}$. Therefore, the Stokes parameters of the laser photons
conserve their values (\ref{20}), (\ref{2.9}).

The Stokes parameters of the final photons are determined with
respect to the transverse to the vector ${\bf k}'$ components of
the polarization 4-vectors $e^{(1)}$ and $(-e^{(2)})$. At small
emission angles $\theta_\gamma \ll 1$, the final photon moves
almost along the direction of the $z$-axis and, therefore, the
components of $e^{(j)}$, which are transverse to the momentum
${\bf k}'$, coincide  within the used accuracy with the vectors
${\bf e}_\perp^{(j)}$ from (\ref{B1}). This means that the Stokes
parameters of the final photons also conserve their forms
(\ref{2.10}), (\ref{2.10f}).

At last, using (\ref{B1}) it is not difficult to check that the
polarization parameters of the electrons $\zeta_j$ and $\zeta_j'$,
defined by (\ref{26}), (\ref{27}) and (\ref{31}), conserve their
forms (\ref{51}) and (\ref{56a}). Therefore, the whole dependence
on $\alpha_0$  enters into the effective cross section and the
polarizations only via the quantity
$x=(4E\omega/m^2)\,\cos^2{(\alpha_0/2)}$.

Finally, one needs to discuss the relation between the direction
of the laser wave linear polarization and the angle $\varphi$
which enters (\ref{zzz}) and the equations in Sect. 3.3. Let us
decompose the vector ${\bf e}_{\rm L}$ along the photon momentum
${\bf k}$ and in the plane orthogonal to that momentum:
 \be
{\bf e}_{\rm L}={\tilde{\bf e}}_{\rm L}+ c{\bf k}\,,\;\;
{\tilde{\bf e}}_{\rm L} {\bf k}=0\,.
 \ee
It is important to note that the electric field of the laser wave
is directed along the vector ${\tilde{\bf e}}_{\rm L}$. Therefore,
this vector ${\tilde{\bf e}}_{\rm L}$ (not the vector ${\bf
e}_{{\rm L}\perp}\,$!) determines the direction of the linear
polarization of the laser photons. To parameterize this direction
we introduce the angle $\gamma$,
 \bea
{\tilde{\bf e}}_{\rm L}&=&{\bf a}\sin{\gamma} + {\bf
b}\cos{\gamma}\,,
 \\
{\bf a}&=& {{\bf k}\times {\bf p}\over |{\bf k}\times {\bf
p}|}\,,\;\; {\bf b}= {{\bf k}\times {\bf a}\over |{\bf k}\times
{\bf a}|}\,,
 \nn
 \\
c &=&- {\sin{\alpha_0}\over (1+\cos{\alpha_0})\omega}
\,\cos{\gamma}\,.
 \nn
 \eea
Here the two unit vectors ${\bf a}$ and ${\bf b}$ are orthogonal
to each other and to the vector ${\bf k}$. Besides, the vector
${\bf a}$ is orthogonal to the colliding plane, defined by the
vectors $\bf p$ and ($-\bf k$), while the vector ${\bf b}$ is in
that plane. The colliding plane should be distinguished from the
scattering plane, defined by the vectors ${\bf p}$ and ${\bf
k}_\perp'$. Let the angle of the scattering plane counted from the
colliding plane be $\varphi^\prime$. Using the above equations it
is not difficult to find that
 \be
{\bf e}_{{\rm L}\perp}={\bf a}\sin{\gamma} - {{\bf k}_\perp\over
|{\bf k}_\perp|}\, \cos{\gamma}\,.
 \ee
This means that the angle $\gamma$ which parameterizes the
direction of the linear polarization is simultaneously the angle
of the vector ${\bf e}_{{\rm L}\perp}$ counted from the colliding
plane. Therefore we conclude that the parameter $\varphi$ is equal
to the difference of two angles, the angle between the two planes
and the angle $\gamma$,
 \be
\varphi=\varphi^\prime -\gamma \, .
 \ee



\begin{thebibliography}{99}
\bibitem{Yokoya}
K.~Yokoya, CAIN2e Users Manual. The latest version is available
from ftp://lcdev.kek.jp/pub/Yokoya/cain235/CainMan235.pdf.
\end{thebibliography}
\end{document}